\documentclass[12pt]{article}
\usepackage{amsmath,amsfonts,amssymb}
\usepackage{mathrsfs}  
\usepackage{graphics}
\usepackage{xcolor}
\usepackage{amsfonts}
\usepackage{hyperref}
\usepackage{times}
\usepackage{bm}
\usepackage{natbib}
\usepackage{booktabs}
\usepackage{amsthm}
\usepackage{tikz}
\usepackage[flushleft]{threeparttable}
\usepackage{longtable}
\usepackage{setspace}
\usepackage{dsfont}
\usetikzlibrary{shapes,decorations,arrows,calc,arrows.meta,fit,positioning}
\tikzset{
    -Latex,auto,node distance =1 cm and 1 cm,semithick,
    state/.style ={ellipse, draw, minimum width = 0.7 cm, fill = yellow!25},
    point/.style = {circle, draw, inner sep=0.04cm,fill,node contents={}},
    bidirected/.style={Latex-Latex,dashed},
    el/.style = {inner sep=2pt, align=left, sloped}
}
\usepackage[ruled,vlined]{algorithm2e}

\makeatletter
\renewcommand{\algocf@captiontext}[2]{#1\algocf@typo. \AlCapFnt{}#2} 
\def\@algocf@capt@plain{top}
\renewcommand{\algocf@makecaption}[2]{%
  \addtolength{\hsize}{\algomargin}%
  \sbox\@tempboxa{\algocf@captiontext{#1}{#2}}%
  \ifdim\wd\@tempboxa >\hsize
    \hskip .5\algomargin%
    \parbox[t]{\hsize}{\algocf@captiontext{#1}{#2}}
  \else%
    \global\@minipagefalse%
    \hbox to\hsize{\box\@tempboxa}
  \fi%
  \addtolength{\hsize}{-\algomargin}%
}
\makeatother

\def\Bka{{\it Biometrika}}

\def\T{{ \mathrm{\scriptscriptstyle T} }}

\def\pr{{ \mathrm{pr} }}

\def\E{{E}}

\newcommand{\lasso}{\mathrm{alasso}}

\newcommand{\aipw}{ {\mathrm{aipw}}}
\newcommand{\acw}{\mathrm{acw}} 
\newcommand{\gbm}{\mathrm{gbm}}
\newcommand{\ppp}{\mathrm{ppp}}
\newcommand{\final}{\mathrm{*}}
\newcommand{\eff}{\mathrm{eff}} 
\newcommand{\bg}{\boldsymbol{g}}
\newcommand{\rct}{\mathcal{R}}
\newcommand{\rwd}{\mathcal{E}}
\def\bV{\mathbb{V}}

\makeatletter
\newcommand*{\addFileDependency}[1]{
  \typeout{(#1)}
  \@addtofilelist{#1}
  \IfFileExists{#1}{}{\typeout{No file #1.}}
}
\makeatother

\newtheorem{lemma}{Lemma}\newtheorem{theorem}{Theorem}
\newtheorem{assumption}{Assumption}\newtheorem{remark}{Remark}

\begin{document}

\def\spacingset#1{\renewcommand{\baselinestretch}%
{#1}\small\normalsize} \spacingset{1}

\newcommand{\blind}{1}
\if1\blind
{
\date{}
  \title{\bf Improving randomized controlled trial analysis via data-adaptive borrowing}
\author{Chenyin Gao$^{1}$, Shu Yang$^{1}$, Mingyang Shan$^{2}$, Wenyu Ye$^{2}$, Ilya Lipkovich$^{2}$, \and Douglas Faries$^{2}$\\
$^{1}$Department of Statistics, North Carolina State University, Raleigh, NC, U.S.A.\\
$^{2}$Eli Lilly\&Company, Indianapolis, IN, U.S.A.
}
  \maketitle
} \fi

\if0\blind
{
  \bigskip
  \bigskip
  \bigskip
  \begin{center}
    {\LARGE\bf  Improving randomized controlled trial analysis via data-adaptive borrowing}
\end{center}
  \medskip
} \fi

\bigskip 
\begin{abstract}
In recent years, real-world external controls have grown in popularity as a tool to empower randomized placebo-controlled trials, particularly in rare diseases or cases where balanced randomization is unethical or impractical. However, as external controls are not always comparable to the trials, direct borrowing without scrutiny may heavily bias the treatment effect estimator. Our paper proposes a data-adaptive integrative framework capable of preventing unknown biases of the external controls. The adaptive nature is achieved by dynamically sorting out a comparable subset of the external controls via bias penalization. Our proposed method can simultaneously achieve (a) the semiparametric efficiency bound when the external controls are comparable and (b) selective borrowing that mitigates the impact of the existence of incomparable external controls. Furthermore, we establish statistical guarantees, including consistency, asymptotic distribution, and inference, providing type-I error control and good power. Extensive simulations and two real-data applications show that the proposed method leads to improved performance over the trial-only estimator across various bias-generating scenarios.
\end{abstract}
\noindent%
\vfill

\doublespacing

\section{Introduction}\label{sec:introduction}

Randomized controlled trials have been considered the gold standard of clinical research to provide confirmatory evidence
on the safety and efficacy of treatments. However, randomized placebo-controlled trials are expensive, require lengthy recruitment periods, and may not always be ethical, feasible, or practical in rare or life-threatening diseases. In response, quality patient-level real-world data from disease registries and electronic
health records have become increasingly available and can generate fit-for-purpose real-world
evidence to facilitate healthcare and regulatory decision-making \citep{fda2021registries}.
Studies using real-world
data may have advantages over randomized placebo-controlled trials including longer observation
windows, larger and more heterogeneous patient populations, and reduced burden on investigators and patients \citep{visvanathan2017untapped,colnet2020causal}.
There is interest in novel clinical trial designs that leverage external controls from real-world data to improve the efficiency of randomized placebo-controlled trials while maintaining robust evidence on the safety and efficacy of treatments \citep{Silverman2018,FDA2019,ghadessi2020roadmap}. The focus of this paper is on hybrid control arm designs using real-world data, where the concurrent control arm is augmented with real-world external controls to form a hybrid comparator group.


The concept of hybrid controls dates back to \citet{pocock1976combination}, which combined the trial data and historical controls by adjusting for the data source level differences. Since then, numerous methods for using the external controls have been developed. However, regulatory approvals of external control arm designs as confirmatory trials are rare and limited to ultra-rare diseases, pediatric trials, or oncology trials \citep{Blinatumomab,Avelumab,Odogwu2018}. Concerns regarding the validity and comparability of the external controls have limited their use in a broader context. Guidance documents from regulatory agencies, including the recent FDA draft guidance \citep{FDA2023}, note several potential issues with the external controls including selection bias, lack of concurrency, differences in the definitions of covariates, treatments, or outcomes, and unmeasured confounding \citep{FDA2001,FDA2019,FDA2023}. Without proper scrutiny, each of these concerns may lead to biased treatment effect estimates and misleading
conclusions.

Selection bias is a type of data heterogeneity often encountered
in non-randomized studies. In the context of external control augmentation, it arises when the real-world baseline subjects' characteristics differ from those in the trial data. Multiple methods are available to adjust for selection bias by balancing the baseline covariates' distributions across the different data sources. For example, matching and subclassification approaches select a subset of comparable external controls to construct the hybrid control arm \citep{stuart2010matching}. Matching on the propensity score or the probability of trial inclusion can balance numerous baseline covariates simultaneously \citep{rosenbaum1983central}. Weighting approaches that re-weight external controls using the probability of trial inclusion or other balancing scores have also been proposed, e.g., empirical likelihood \citep{qin2015using}, entropy balancing \citep{lee2021improving, chu2023targeted, wu2022transfer}, constrained maximum likelihood \citep{chatterjee2016constrained,zhang2020generalized}, and Bayesian power priors \citep{neuenschwander2010summarizing,van2018including}. Furthermore, matching or weighting can be combined with the outcome modeling to enhance robustness against model misspecification in addressing selection bias of external controls \citep{li2021improving}.

Differences in the outcomes may still exist between the concurrent controls
and the external controls after matching or weighting due to differences
in study settings, time frame, data quality, or definition of covariates
or outcomes \citep{phelan2017illustrating}. Methods were proposed to adaptively select the degree of borrowing or adjust the outcomes for external controls based on the observed outcome differences with the
concurrent controls. Some researchers suggested first testing the heterogeneity in control outcomes before deciding whether to incorporate external subjects into the hybrid control arm \citep{viele2014use, li2021improving}. More dynamic
borrowing approaches were also proposed including matching and bias adjustment \citep{stuart2008matching}, power priors \citep{Ibrahim2000,Neuenschwander2009}, Bayesian hierarchical models including meta-analytic predictive priors \citep{neuenschwander2010summarizing,schoenfeld2019design}, and commensurate priors \citep{Hobbs2011}. While these existing methods seem appealing, simulation studies could not identify a single approach that could perform well across all scenarios where hidden biases exist \citep{Shan2022}. The surveyed Bayesian methods often have inflated type I errors while Frequentist methods suffer lower power when hidden biases exist. Nearly all methods performed poorly in the presence of unmeasured confounding and could not simultaneously minimize bias and gain power. Further, many existing methods rely on parametric assumptions that are sensitive to model misspecification and cannot capture complex relationships that are prevalent in practice.

In this paper, we propose an approach to achieve an efficient estimation of treatment effects that is robust to various potential discrepancies that may arise in the external controls. When handling the selection bias of the external controls, our proposal is based on calibration weighting \citep{lee2021improving} so that the covariate distribution of external controls matches with that of the trial subjects. Furthermore, leveraging semiparametric theory, we develop an integrative augmented calibration weighting estimator, motivated by the efficient influence function (\citealp{bickel1993efficient,tsiatis2007semiparametric}),
which is semiparametrically efficient and doubly robust against model
misspecification. Despite the potential to view the selection bias problem as a generalizability or transportability issue \citep{lee2021improving},
our framework fundamentally diverges from theirs as our context encompasses the outcomes from both the trial data and external controls, while
Lee et al. solely considered the trial outcomes.

To deal with potential outcome heterogeneity, we develop a selective borrowing framework to determine an optimal subset from the external controls for integration. Specifically, we introduce a bias parameter for each external subject entailing his or her comparability with the concurrent control. To prevent bias in the integrative estimator, the goal is to select the comparable external controls with zero bias and exclude any others with non-zero bias. Thus, this formulation recasts the selective borrowing strategy as a model selection problem, which can be solved by penalized estimation (e.g., the adaptive lasso penalty; \citealp{zou2006adaptive}). Subsequent to the selection process, comparable external controls are utilized to construct the integrative estimator. Prior works such as those by \cite{chen2021combining}, \cite{liu2021propensity}, and \cite{zhai2022data} although able to identify biases, exclude the entire external sample when confronted with incomparability. Moreover, compared to these existing selective borrowing
approaches, our method leverages off-the-shelf machine learning models
to achieve semiparametric efficiency and does not require strigent parametric assumptions on the distribution of outcomes. 


\section{Methodology \label{sec:Methodology}}

\subsection{Notation, assumptions, and objectives \label{sec:notations}}

Let $\rct$ represent a randomized placebo-controlled trial and $\rwd$
represent an external control source, which contains $N_{\rct}$
and $N_{\rwd}$ subjects, respectively. The total sample size
is $N=N_{\rct}+N_{\rwd}$. An extension to multiple
external control groups is discussed in $\S$\ref{sec:Extension} of the Supplementary Material. A total of $N_{t}$ and $N_{c}$ subjects receive the active treatment and control treatment in $\rct$, while we assume all $N_{\rwd}$
subjects in $\rwd$ receive the control. Each observation
$i\in\rct$ comprise the outcomes $Y_{i}$, the treatment assignment
$A_{i}$, and a set of baseline covariates $X_{i}$. Similarly, each
observation $i\in\rwd$ comprise $Y_{i}$, $A_{i}$, and $X_{i}$.
Let $R_{i}$ represent a data source indicator, which is $1$ for
all subjects $i\in\rct$ and is $0$ for all subjects $i\in\rwd$.
To sum up, an independent and identically distributed sample $\{V_{i}:i\in\rct\cup\rwd\}$
is observed, where $V=(X,A,Y,R)$. Let $Y(a)$ denote the potential
outcomes under treatment $a$ \citep{rubin1974estimating}. The causal
estimand of interest is defined as the average treatment effect among
the trial population, $\tau=\mu_{1}-\mu_{0}$, where $\mu_{a}=\E\{Y(a)\mid R=1\}$
for $a=0,1$. The clinical trials for treatment effect estimation satisfy the following assumption.

\begin{assumption}[Consistency, randomization and positivity]\label{a:ign} 
(i) $Y=AY(1)+(1-A)Y(0)$, (ii) $Y(a)\perp\!\!\!\perp A\mid(X,R=1)$ for $a=0,1,$ and (ii) the
known treatment propensity score satisfies that $1>\pi_{A}(x)=\pr(A=1\mid X=x,R=1)>0$ for all $x$ such that $\pr(X=x,R=1)>0$. 
\end{assumption}

Assumption \ref{a:ign} is standard in the causal inference literature
\citep{rosenbaum1983central,imbens2004nonparametric}
and holds for the well-controlled clinical trials guaranteed
by the randomization mechanism. Under Assumption \ref{a:ign}, $\tau$
is identifiable with the trial data. 

Moreover, the external controls should ideally be comparable with the concurrent controls.

\begin{assumption}[External control compatibility]\label{assum:exchange_delta}
(i) $\E\left\{ Y(0)\mid X=x,R=0\right\} =\E\{Y(0)\mid X=x,R=1\}$, and (ii) $\pr(R=1\mid X=x)>0$ for all $x$ such that $\pr(X=x,R=0)>0$. 
\end{assumption}

Assumption \ref{assum:exchange_delta} states that the conditional mean of $Y(0)$ is the same for the trial data and external controls. This assumption holds if $X$ captures all the outcome predictors that are correlated with $R$. From the guidance in \cite{FDA2023} for drug development in rare diseases, there
are five main concerns regarding the use of external controls: (i) selection bias, (ii) unmeasured confounding, (iii) lack of concurrency, (iv) data quality, and (v) outcome validity. Assumption \ref{assum:exchange_delta} does not require the covariate distribution of external controls to be the same as that of the trial data, which is termed as the selection bias in the guidance. Under Assumption \ref{assum:exchange_delta}, borrowing external controls to improve treatment effect estimation is similar to a transportability or covariate shift problem. However, the presence of concerns (ii)–(v)
can result in violation of Assumption \ref{assum:exchange_delta}.
Our paper has two main objectives: 1) Under Assumption \ref{assum:exchange_delta}, similarly to the work of \cite{li2021improving}, we develop a semiparametrically efficient and robust strategy to borrow
external controls to improve estimation while correcting for selection bias
($\S$\ref{subsec:Semiparametric-efficient-estimat}); 2) Considering that Assumption \ref{assum:exchange_delta} can be potentially violated, we incorporate a selective borrowing procedure that will detect the biases and retain only a subset of comparable external controls for integration ($\S$\ref{subsec:selective borrowing}). 

\subsection{Semiparametric efficient estimation under the ideal situation\label{subsec:Semiparametric-efficient-estimat}}

From the semiparametric theory \citep{bickel1993efficient},
we derive efficient and robust estimators for $\tau$ under Assumptions
\ref{a:ign} and \ref{assum:exchange_delta}. The derivation reaches the same estimator as \cite{li2021improving}, and will serve as the base for our selective borrowing strategy. The semiparametric model
is attractive as it exploits the observed data without making assumptions about the nuisance parts of the data generation process that are not of substantive interest. We derive the efficient influence function of $\tau$ in Theorem \ref{thm:tau_EIF}, which shall serve as the foundational component of our proposed framework.
\begin{theorem}\label{thm:tau_EIF} Under Assumptions \ref{a:ign} and \ref{assum:exchange_delta}, the efficient influence function of $\tau$ is 
\begin{multline}
 \psi_{\tau,\eff}(V;\mu_{1},\mu_{0},q,r)=\frac{R}{\pr(R=1)}\left[\{\mu_{1}(X)-\mu_{0}(X)-\tau\}+\frac{A\{Y-\mu_{1}(X)\}}{\pi_{A}(X)}\right] \\
  -\frac{R(1-A)+(1-R)r(X)}{\pr(R=1)}\frac{q(X)\left\{ Y-\mu_{0}(X)\right\}}{q(X)\{1-\pi_{A}(X)\}+r(X)}, 
    \label{eq:EIF_tau_c}
\end{multline}
where $\mu_{1}(X)=\E(Y\mid X,R=1,A=1)$, $\mu_0(X)=\E(Y\mid X, R=1, A=0)=\E(Y\mid X, R=0)$, $r(X)=\text{var}(Y\mid X,R=1, A=0)/\text{var}(Y\mid X,R=0)$, and $q(X)=\pr(R=1\mid X)/\pr(R=0\mid X)$. 
\end{theorem}
Based on Theorem \ref{thm:tau_EIF}, the semiparametric efficiency
bound for $\tau$ is $\bV_{\tau,\eff}=\E\{\psi_{\tau,\eff}^{2}(V;\mu_{1},\mu_{0}$, $q,r)\}$.
Hence, a principled estimator can be motivated by solving the empirical analog of $\E\{\psi_{\tau,\eff}(V$ $;\mu_{1},\mu_{0},q,r)\}=0$ for $\tau$.

Let the estimators of $(\mu_{0},\mu_{1},q,$$r)$ be $(\widehat{\mu}_{0},\widehat{\mu}_{1},\widehat{q},\widehat{r})$,
and denote $\widehat{\epsilon}_{a,i}=Y_{i}-\widehat{\mu}_{a}(X_{i})$
$(a=0,1)$. Then, by solving the empirical version of the efficient influence function for $\tau$, we have 
\begin{multline}
\widehat{\tau}  =\frac{1}{N_{\rct}}\sum_{i\in\rct\cup\rwd}R_{i}\left\{\widehat{\mu}_{1}(X_{i})-\widehat{\mu}_{0}(X_{i})+\frac{A_{i}\widehat{\epsilon}_{1,i}}{\pi_{A}(X_{i})}\right\} \\
  -\frac{1}{N_{\rct}}\sum_{i\in\rct\cup\rwd}\frac{\{R_{i}(1-A_{i})+(1-R_{i})\widehat{r}_{i}(X_{i})\}\widehat{q}(X_{i})}{\widehat{q}(X_{i})\{1-\pi_{A}(X_{i})\}+\widehat{r}(X_{i})}\widehat{\epsilon}_{0,i}.
  \label{eq:ACW}
\end{multline}
We now discuss the estimators for the nuisance functions $(\mu_{0},\mu_{1},q,$$r)$.
To estimate $\mu_{0}(X)$, $\mu_{1}(X)$, and $r(X)$, one can follow
the standard approach by fitting parametric models based
on the trial data. 

For estimating the weight $q(X)$, a direct approach
is to predict $\pr(R=0\mid X)$, which however is unstable due to inverting probability estimates. To achieve stability of weighting,
the key insight is based on the central role of $q(X)$ as balancing
the covariate distribution between two groups: $\E\{(1-R)q(X)\bg(X)\}=\E\{R\bg(X)\}$ for any $\bg(X)=\{g_{1}(X),\ldots,g_{K}(X)\}$, which is a $K$-dimensional function of $X$. Thus, we estimate $q(X)$ by calibrating the covariate balance between the trial data and external controls. In particular, we assign a weight $q_{i}$ for each subject $i\in \rwd$, then solve the following optimization problem
for $Q=\{q_{i}:i\in\rwd\}$: $$\min_{q}L(Q)=\sum_{i\in\rwd}q_{i}\log q_{i},$$ subject to (i) $q_{i}\ge0$, $i\in\rwd,$ (ii) $\sum_{i\in\rwd}q_{i}\bg(X_{i})=\sum_{i\in\rct}\bg(X_{i})$.
First, $L(Q)$ is the entropy of the weights;
thus, minimizing this criterion ensures that the calibration weights
are not too far from uniform so it minimizes the variability due to
heterogeneous weights. Constraint (i) is a standard condition for the weights. Constraint (ii) forces the empirical moments
of the covariates to be the same after calibration, leading to better-matched distributions of the trial data and external controls. 

The optimization problem can be solved using constrained convex optimization. The estimated calibration weight is $\widehat{q}_{i}=q(X_{i};\widehat{\eta})=\exp\{\widehat{\eta}^{\top}\bg(X_{i})\}$, and $\widehat{\eta}$ solves 
$$U(\eta)=\sum_{i\in\mathcal{\rwd}}\exp\left\{ \eta^{\top}\bg(X_{i})\right\} \bg(X_{i})-\sum_{i\in\rct}\bg(X_{i}) =0,$$
which is the Lagrangian dual problem to the optimization problem. The dual problem
also entails that the calibration weighting approach makes a log regression
model for $q(X)$. We term $\widehat{\tau}$ with the calibration
weights by the augmented calibration weighting estimator $\widehat{\tau}_{\acw}$. 

\begin{remark}\label{rmk:variance_ratio}
The variance ratio $r(X)$ quantifies
the relative residual variability of $Y(0)$ given $X$ between the
trial data and external controls. In general, estimating the conditional variance ratio involves nonparametric regression, which can be challenging; see \cite{shen2020optimal} and references therein. Fortunately, the consistency of $\widehat{\tau}_{\acw}$
does not rely on the correct specification of $r(X)$. For example,
if $\widehat{r}(X)$ is set to be zero, $\widehat{\tau}_{\acw}$ reduces
to the trial-only estimator without borrowing any external information, which is always consistent. In order to leverage external information and estimate
$r(X)$ practically, we can make a simplifying homoscedasticity assumption
that the residual variances of $Y(0)$ after addressing $X$ are constant over studies. In this case, $r(X)$ can be estimated by $\widehat{r}=N_{\rwd}N_{c}^{-1}\sum_{i\in\rct}(1-A_{i})\{Y_{i}-\widehat{\mu}_{0}(X_{i})\}^{2}/\sum_{i\in\rwd}\{Y_{i}-\widehat{\mu}_{0}(X_{i})\}^{2}$.
\end{remark}

We show that $\widehat{\tau}_{\acw}$ has the following
desirable properties: 1) local efficiency, $\widehat{\tau}_{\acw}$
achieves the semiparametric efficiency bound if the nuisance functions are correctly specified; 2) double robustness, $\widehat{\tau}_{\acw}$
is consistent for $\tau$ if either the model for $\mu_{a}(X)$ or that for $q(X)$ is correct; see proof in the $\S$\ref{subsec:proof_dr_acw} of the Supplementary Material.

The doubly robust estimators were initially developed to gain robustness
to parametric misspecification but are now known to also be robust
to approximation errors using machine learning methods \citep[e.g.,][]{chernozhukov2018double}.
We will investigate this new doubly robust feature for the proposed
estimator $\widehat{\tau}_{\acw}$, and use flexible semiparametric or nonparametric methods to estimate both $\mu_{a}(X)$
($a=0,1$), $r(X)$ and $q(X)$ in (\ref{eq:ACW}). First, we will consider the method
of sieves \citep{chen2007large} for $q(X)$. In comparison with other
nonparametric methods such as kernels, the method of sieves is particularly
well-suited for calibration weighting. We consider general sieve basis
functions such as power series, Fourier series, splines, wavelets,
and artificial neural networks; see \citet{chen2007large} for a comprehensive
review. The number of bases can be selected by cross validation. Second, we consider flexible outcome models, e.g., generalized additive models, kernel regression, and the method of sieves for $\mu_{a}(X)$ ($a=0,1$). Using flexible methods alleviates bias from the misspecification of parametric models.
The following regularity conditions are required for the nuisance
function estimators.
\begin{assumption}\label{assum:rate-conditions} For a function $f(X)$
with a generic random variable $X$, define its $L_{2}$-norm as $\|f(X)\|=\{\int f(x)^{2}d\pr(x)\}^{1/2}$.
Assume: (i) $\|\widehat{\mu}_{a}(X)-\mu_{a}(X)\|=o_{p}(1),a=0,1$ and $\|\widehat{q}(X)-q(X)\|=o_{p}(1)$; (ii) $\|\widehat{\mu}_{0}(X)-\mu_{0}(X)\|\times\|\widehat{q}(X)-q(X)\|=o_{p}(N^{-1/2})$; (iii) $\|\widehat{r}(X)-r^*(X)\|=o_{p}(1)$ for some $r^*(X)$, and (iv) additional regularity conditions in Assumption \ref{assum:regularity-eff}
of the Supplementary Material.

\end{assumption}

Assumption \ref{assum:rate-conditions} is a set of typical regularity
conditions for M-estimation to achieve rate double robustness \citep{van2000asymptotic}.
Under these regularity conditions, our proposed framework can incorporate
flexible methods for estimating the nuisance functions while remains
the parametric-rate consistency for $\widehat{\tau}_{\acw}$. 

\begin{theorem}

\label{thm:rate_robustness} Under Assumptions \ref{a:ign}-\ref{assum:rate-conditions}, we have $N^{1/2}(\widehat{\tau}_{\acw}-\tau)\overset{d}{\rightarrow} N(0,\bV_{\tau})$,
where $\bV_{\tau}=\E\{\psi_{\tau,\eff}^{2}(V;\mu_{1},\mu_{0},q,r^{*})\}$. If $r^*(X)=r(X)$, $\widehat{\tau}_{\acw}$
achieves the semiparametric efficiency. 
\end{theorem}
Theorem \ref{thm:rate_robustness} motivates variance estimation by $\widehat{\bV}_r=N^{-1}\sum_{i\in\rct\cup\rwd}\psi^2_{\widehat{\tau}_{\acw},\eff}(V_i;\widehat{\mu}_1,\widehat{\mu}_0,\widehat{q},\widehat{r})$, which is consistent for $V_\tau$ under Assumptions \ref{a:ign}-\ref{assum:rate-conditions}. 

\subsection{Bias detection and selective borrowing \label{subsec:selective borrowing}}

In practical situations, Assumption \ref{assum:exchange_delta} may
not hold, and the augmentation in (\ref{eq:ACW}) can be biased.
We will develop a selective borrowing framework to select external subjects that are comparable with the concurrent controls for integration. To account for the potential violations,
we introduce a vector of bias parameter $b_{0}=(b_{1,0},\ldots,b_{N_{\rwd},0})$ for
all $i\in\rwd$, where $b_{i,0}=b_0(X_i)=\E(Y_{i}\mid X_{i},A_i=0, R_{i}=0)-\E(Y_{i}\mid X_{i},A_i=0, R_{i}=1)=\mu_{0,\rwd}(X_{i})-\mu_{0}(X_{i})$. When Assumption \ref{assum:exchange_delta}
holds, we have $b_{0}=0$. Otherwise, there
exists at least one $i\in\rwd$ such that $b_{i,0}\neq0$.
To prevent bias in $\widehat{\tau}_{\acw}$ from incomparable external controls, the goal is to select the comparable subset with $b_{i,0}=0$ and exclude any others with $b_{i,0}\neq0$. 

Let $\widehat{b}_{i}=\widehat{\mu}_{0,\rwd}(X_{i})-\widehat{\mu}_{0}(X_{i})$
be a consistent estimator for $b_{i,0},$ where $\widehat{\mu}_{0,\rwd}(X_{i})$
is a consistent estimator for $\mu_{0,\rwd}(X_{i})$. Let $\widehat{b}=(\widehat{b}_{1},\ldots,\widehat{b}_{N_{\rwd}})$
be an initial estimator for $b_{0}$. We propose a refined estimator of $b_{0}$ by penalized estimation:
\begin{equation}
\tilde{b}=\arg\min_{b}\left\{ (\widehat{b}-b)^{\T}\widehat{\Sigma}_{b}^{-1}(\widehat{b}-b)+\lambda_{N}\sum_{i\in\rwd}p(|b_{i}|)\right\} ,\label{eq:penalized_log_likelihood:}
\end{equation}
where $\widehat{\Sigma}_{b}$ is the estimated variance of $\widehat{b}$, $p(|b_{i}|)=|b_{i}|/|\widehat{b}_{i}|^{\nu}$
is the adaptive lasso penalty term, and $(\lambda_{N},\nu)$ are
two tuning parameters.
Intuitively, if $\widehat{b}_{i}$ is close to zero, the associated
penalty will be large, which further shrinks the estimate $\tilde{b}_{i}$
towards zero. According to \citet{zou2006adaptive}, \cite{huang2008adaptive}, and \cite{lin2009adaptive}, the adaptive lasso penalty can lead to a desirable property under the following regularity conditions.
\begin{assumption}\label{assum:selection-regularity}
(i) $a_N\max_i\{\widehat{\mu}_{0}(X_i)- \mu_{0}(X_i)\}=O_p(1)$ and $a_N\max_i\{\widehat{\mu}_{0,\rwd}(X_i)- \mu_{0,\mathcal{E}}(X_i)\}=O_p(1)$, $\forall i\in \mathcal{E}$; (ii) Let $\tau_{b,\min}$ and $\tau_{b,\max}$ be the smallest and largest eigenvalues of $\widehat{\Sigma}_b$, there exist constants $\tau_1$ and $\tau_2$ such that $0<\tau_1\leq \tau_{b,\min} \leq \tau_{b,\max} \leq \tau_2$; (iii) $a_N b_{\min}\rightarrow\infty$, where $b_{\min}=\min\{b_{i,0}, i\notin\mathcal{A}\}$, and (iv) $\lambda_N / b_{\min}^{\nu+1}\rightarrow 0$ and $\lambda_N a_N^\nu \rightarrow \infty$.
\end{assumption}
\begin{lemma}
\label{lem:selection-consistency} Suppose the assumptions in Theorem
\ref{thm:rate_robustness} and Assumption \ref{assum:selection-regularity} hold except that Assumption \ref{assum:exchange_delta}
may be violated, we have $\lim_{N\rightarrow\infty}\pr(\tilde{\mathcal{A}}=\mathcal{A})=1$.
\end{lemma}
Lemma \ref{lem:selection-consistency} shows that the adaptive lasso
penalty has the ability to select zero-valued parameters consistently
when using a $a_N$-consistent initial estimator $\widehat{b}_i$ and proper choices of $(\lambda_{N},\nu)$, provided that the minimum of the non-zero bias $b_{\min}$ does not diminish too fast and the initial estimator $\widehat{b}_i$ is sufficient good. In practice, the initial estimator $\widehat{b}_i$ can be obtained by leveraging off-the-shelf machine learning models with guaranteed convergence rate, and $(\lambda_{N},\nu)$ are selected by minimizing the mean square error using cross validation. Given $\tilde{b}$, the selected set of comparable external controls is $\tilde{\mathcal{A}}=\{i:\tilde{b}_{i}=0\}$. The modified integrative estimator is 
\begin{multline}
\widehat{\tau}_{\acw}^{\lasso}=  \frac{1}{N_{\rct}}\sum_{i\in\mathcal{\rct}\cup\rwd}R_{i}\left[\widehat{\mu}_{1}(X_{i})-\widehat{\mu}_{0}(X_{i})+\frac{A_{i}\widehat{\epsilon}_{1,i}}{\pi_{A}(X_{i})}\right] \\
  -\frac{1}{N_{\rct}}\sum_{i\in\mathcal{\rct}\cup\rwd}\frac{\{R_{i}(1-A_{i})+(1-R_{i})\widehat{r}_{b}(X_{i})\mathds{1}(\tilde{b}_{i}=0)\}\widehat{q}(X_{i})}{\widehat{q}(X_{i})\{1-\pi_{A}(X_{i})\}+ \widehat{r}_{b}(X_i)\pr(\tilde{b}_{i}=0\mid X,R=0)}\widehat{\epsilon}_{0,i},\label{eq:aug-data-dep}
\end{multline}
where $\widehat{r}_{b}(X)$ is the estimated function of $r_{b}(X)=\text{var}(Y\mid X,R=1, A=0)/\text{var}(Y\mid X,R=0,b_{0}=0)$, which are used to adjust for the changes in the covariate distribution
from all external controls in $\rwd$ to $\tilde{\mathcal{A}}$.

Followed by the suggestions in \citet{ho2007matching} to improve the finite-sample performances, the nearest-neighbor matching based on the estimated probability of trial inclusion $e(X)=\pr(R=1\mid X)$ is performed after selecting the comparable subset $\tilde{\mathcal{A}}$, which is to ensure a more balanced allocation ratio between the treated group and the hybrid control arm; see Algorithm \ref{alg:flowchart} for the overview of our selective borrowing framework.
\begin{algorithm}[!htbp]
\caption{\label{alg:flowchart} the proposed selective integrative estimator}
{Input}: A randomized controlled trial with size $N_{\rct}=N_t+N_c$ and external controls. \\
{Step 1}: fit the models for the outcome means $\mu_1, \mu_0, \mu_{0,\rwd}$, and the weights $q$.\\
{Step 2}: construct the pseudo-observation $\widehat{\xi}$ for the bias parameter $b_0$.\\
{Step 3}: select the comparable subset $\tilde{\mathcal{A}}=\{i:\tilde{b}_i=0\}$ via the bias penalization (\ref{eq:penalized_log_likelihood:}).\\
{Step 4}: if $|\tilde{\mathcal{A}}|>N_t-N_c$, 
perform the nearest-neighbor matching to select $N_t-N_c$ external controls as the final $\tilde{\mathcal{A}}$;
else, {jump to Step 5.}\\
{Step 5}: compute $\widehat{\tau}_{\acw}^{\lasso}$ in (\ref{eq:aug-data-dep}) using the selected external controls in $\tilde{\mathcal{A}}$.
\end{algorithm}

We show the efficiency gain of the proposed estimator compared to the trial-only estimator.

\begin{theorem}\label{thm:acw-final-point}

Suppose the assumptions in Theorem
\ref{thm:rate_robustness} and Assumption \ref{assum:selection-regularity} hold except that Assumption \ref{assum:exchange_delta}
may be violated.
Let $r_{b}^{*}(X)=r_b(X)$, the reduction of the asymptotic variance of $\widehat{\tau}_{\acw}^{\lasso}$ compared to the trial-only estimator is: 
\begin{align}
 & \frac{1}{\pr^{2}(R=1)}\E\left[\frac{\pr(R=1\mid X)r_b(X)\mathds{1}(b_{0}=0)\text{var}(Y\mid X,R=1, A=0)}{
 \{q(X)\{1-\pi_{A}(X)\} + r_b(X)\pr(b_{0}=0\mid X, R=0)\}
 \{1-\pi_A(X)\}}
\right],\label{eq:efficiency-gain}
\end{align}
which is strictly positive unless $r_{b}(x)=0$ or $b_{0}\neq 0$ or $\text{var}(Y\mid X,R=1,A=0)=0$ for all $x$ such that $\pr(X=x)>0$. 
\end{theorem}

We derive (\ref{eq:efficiency-gain}) using orthogonality of the efficient influence function of $\tau$ to the nuisance tangent space, and relegate the details
to the Supplemental Material. Theorem \ref{thm:acw-final-point} showcases
the advantage of including external controls in a data-adaptive manner,
where the asymptotic variance of $\widehat{\tau}_{\acw}^{\lasso}$
should be strictly smaller than the trial-only estimator unless the
external controls all suffer exceeding noises, i.e., $r_{b}(X_{i})=0$ or the compatible subset $\mathcal{A}$ of the external controls is an empty set, i.e., $b_0\neq 0$ or the covariate $X$ captures all the variability of $Y(0)$ in the trial data, i.e., $\text{var}(Y\mid X,R=1,A=0)=0$. Below, we establish the asymptotic
properties and provide a valid inferential framework for the proposed
integrative estimator; more details are provided in $\S$\ref{subsec:proofCI} of the Supplemental Material.

\begin{theorem}\label{thm:acw-final-CI}
Suppose the assumptions in Theorem
\ref{thm:rate_robustness} and Assumption \ref{assum:selection-regularity} hold except that Assumption \ref{assum:exchange_delta}
may be violated,
we have $N^{1/2}(\widehat{\tau}_{\acw}^{\lasso}-\tau)\rightarrow N(0,\bV_{\tau}^{\lasso})$.
Further, the $(1-\alpha)\times100\%$ confidence interval $[L_{\tau},U_{\tau}]$
for $\tau$ can be constructed by 
\begin{equation}
[L_{\tau},U_{\tau}]=\left[\widehat{\tau}_{\acw}^{\lasso}-z_{\alpha/2}\sqrt{\widehat{\bV}_{\tau}^{\lasso}/N},\widehat{\tau}_{\acw}^{\lasso}+z_{\alpha/2}\sqrt{\widehat{\bV}_{\tau}^{\lasso}/N}\right],\label{eq:CI_acw_lasso}
\end{equation}
where $\widehat{\bV}_{\tau}^{\lasso}$ is a variance estimator of $\bV_{\tau}^{\lasso}$,
$z_{\alpha/2}$ is the $1-\alpha/2$ quantile for the standard
normal distribution, and $[L_{\tau},U_{\tau}]$ satisfies
that $\pr(\tau\in[L_{\tau},U_{\tau}])\rightarrow1-\alpha$
as $N\rightarrow\infty$. 

\end{theorem}

\section{Simulation \label{sec:simulation}}

In this section, we evaluate the finite-sample performance of the
proposed framework to estimate treatment effects under potential bias scenarios via plasmode simulations. First, a set of $d=12$ baseline
covariates $X\in\mathbb{R}^d$ is generated by mimicking the correlation structure
and the moments (up to the sixth) of variables from an oncology randomized placebo-controlled trial (i.e., the trial data) and the Flatiron Health Spotlight Phase 2 cohort (Copyright©2020 Flatiron Health, Inc. All Rights Reserved; external controls). 

Next, we generate the data source indicator $R_{i}$ as $R_{i}\mid X_{i},U_{i}\sim\text{Bernoulli}\{\pi_{R}(X_{i},U_{i})\}$
given the sample sizes $(N_{\rct},N_{\rwd})$, where
$U_{i}$ represents an unmeasured confounder. The treatment assignment for the
trial data is completely at random (i.e., $\pr(A_{i}=1\mid R_{i}=1)=N_{t}/N_{\rct}$), while all external subjects receive the control (i.e., $\pr(A_{i}=0\mid R_{i}=0)=1$).
The outcomes $Y_{i}$ are generated
by 
\begin{align*}
&Y_{i}\mid (X_{i},A_{i},U_{i},R_{i}=1)\sim N\{\mu_{0}(X_{i},U_{i},A_{i}),\sigma_{Y}^{2}\}, \\
&Y_{i}\mid (X_{i},U_{i},R_{i}=0)\sim N\{\mu_{0,\rwd}(X_{i},U_{i}),\sigma_{Y}^{2}\}.
\end{align*}
We consider three data-generating scenarios in Table \ref{tab:model-choice-sim}(a), where $\eta_0$ is chosen adaptively to ensure the desired sample sizes $(N_{\rct}, N_{\rwd})$, and $(\eta, \beta, \tilde{\eta}, \tilde{\beta}, \sigma_{Y}^{2})$ are chosen empirically based on the model fits using the observed oncology clinical trial data. In all the scenarios, we use the linear predictor of $X$ to fit $(q,\mu_0,\mu_{0,\rwd})$, and thus the models are correctly specified under the model choices `C' where the linear predictor of $X$ governs the true data generation, but are misspecified under
the choices `W', where the data generation depends on a new set of covariates $\tilde{X}$, which include the quadratic and cubic terms of the $(d-1)$-th and $d$-th covariates (i.e., $X_{d-1}^2, X_d^2, X_{d-1}^3, X_{d}^3$) addition to the baseline covariate $X$. Moreover, we utilize the cross-fitting procedure to select tuning parameters for the gradient boosting model.

\begin{table}[htbp]
\caption{Simulation settings: (a) model choices (C and W), where $\tilde{X}=[X,X_{d-1}^{2},X_{d}^{2},X_{d-1}^{3},X_{d}^{3}]$, and (b) the description of five estimators}
\vspace{.5em}
\resizebox{\textwidth}{!}{%
\begin{tabular}{lccc}
\toprule
(a) & $\text{logit}\{\pi_{R}(X,U)\}$ & $\mu_{0}(X,U,A)$ & $\mu_{0,\rwd}(X,U)$\\
\midrule
C & $\eta_{0}+\eta^{\T}X+\omega U$ & $\beta^{\T}X+A\alpha^{\T}(1, X)+\omega U\sigma_{Y}$ & $\beta^{\T}X+\omega U\sigma_{Y}+ \omega \sigma_Y$\\
W & $\eta_{0}+\tilde{\eta}^{\T}\tilde{X}+\omega U $ & $\tilde{\beta}^{\T}\tilde{X}+A\alpha^{\T}(1,X)+\omega U\sigma_{Y}$ & $\tilde{\beta}^{\T}\tilde{X}+\omega U\sigma_{Y} + \omega \sigma_Y$\\
\bottomrule
\toprule
\multicolumn{4}{l}{(b) Estimators}\\
\midrule
$\widehat{\tau}_{\aipw}$ & \multicolumn{3}{l}{the augmented inverse probability weighting estimator without borrowing \citep{cao2009improving}}       \\
$\widehat{\tau}_{\acw}$        & \multicolumn{3}{l}{the integrative augmented calibration weighting estimator with full borrowing \citep{li2021improving}}       \\
$\widehat{\tau}_{\acw}^{\lasso}$  & \multicolumn{3}{l}{the data-adaptive integrative estimator using the linear regressions for $(\mu_{0},\mu_{0,\rwd})$}       \\
$\widehat{\tau}_{\acw,\gbm}^{\lasso}$  & \multicolumn{3}{l}{the data-adaptive integrative estimator using the tree-based gradient boosting for $(\mu_{0},\mu_{0,\rwd})$}       \\
$\widehat{\tau}_{\ppp}$  & \multicolumn{3}{l}{the Bayesian predictive p-value power prior estimator \citep{kwiatkowski2023case}}\\
\bottomrule
\end{tabular}}\label{tab:model-choice-sim}
\end{table}

The proposed framework is evaluated on an imbalanced trial data where $N_{c}=(20,30,40,50,75$, $100)$ and $N_{t}=200$ with an external control group of size $N_{\rwd}=3000$. We investigate the performance of our proposed estimator under two levels of unmeasured confounding ($\omega=0$
and $0.3$) by comparing with other estimators in Table \ref{tab:model-choice-sim}(b). The trial-only augmented inverse probability weighting estimator, $\widehat{\tau}_{\aipw}$ \citep{cao2009improving}, and the augmented calibration weighting estimator $\widehat{\tau}_{\acw}$ with full borrowing \citep{li2021improving} are used as benchmarks. Two data-adaptive integrative estimators, $\widehat{\tau}_{\acw}^{\lasso}$ and $\widehat{\tau}_{\acw,\gbm}^{\lasso}$, are considered, where linear regressions and tree-based gradient boosting are used to estimate the nuisance models. Other machine learning algorithms that satisfy pointwise consistency, such as the generalized additive model, can also be utilized to select a comparable subset of external controls consistently. The Bayesian predictive p-value power prior estimator, $\widehat{\tau}_{\ppp}$, is an extension of the power prior, which discounts each external control according to its outcome compatibility using Box's p-value \citep{kwiatkowski2023case}.


\begin{figure}[htbp]
    \centering
    \includegraphics[width =0.85\linewidth]{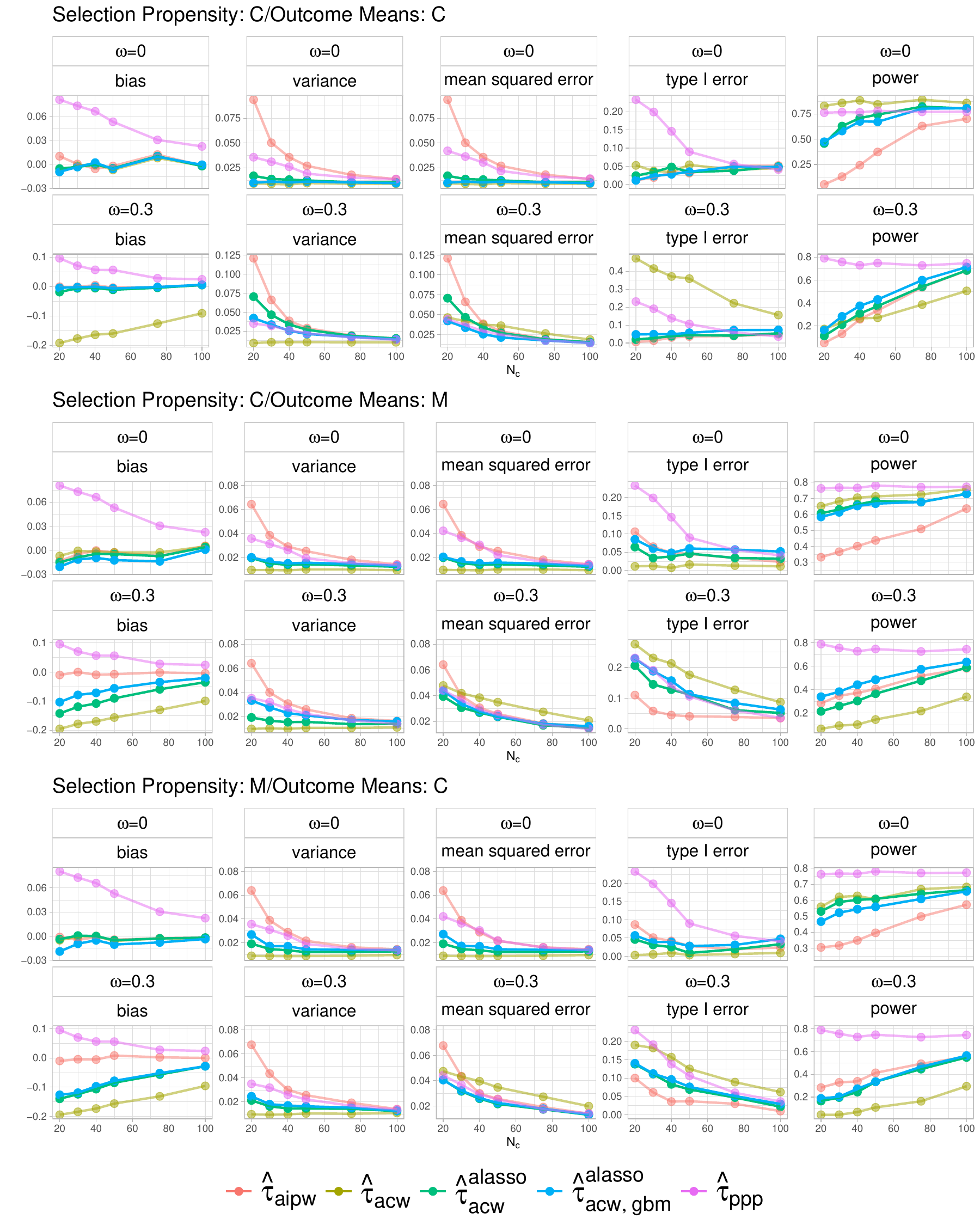}
\caption{\label{fig:correct_modeling-all} Simulation results under various
levels of $\omega$, and different model choices of $q(X)$ and $\mu_{0}(X)$.}
\end{figure}

Figure \ref{fig:correct_modeling-all} displays the average bias,
variance, mean squared error, type I error when $\E\{\tau(X)\mid R=1\}=0$,
and power for testing $\tau>0$ when $\E\{\tau(X)\mid R=1\}=0.3$ based
on $1000$ sets of data replications. Over the three model scenarios, the trial-only estimator $\widehat{\tau}_{\aipw}$
is always consistent but lacks efficiency as it only utilizes the concurrent controls for estimation, especially when $N_c$ is small. When the conditional mean exchangeability
in Assumption \ref{assum:exchange_delta} holds (i.e., $\omega=0$),
the full-borrowing estimator $\widehat{\tau}_{\acw}$ is most efficient,
shown by its low mean squared error and high power for detecting a significant treatment
effect. Our proposed selective integrative estimators, $\widehat{\tau}_{\acw}^{\lasso}$ and $\widehat{\tau}_{\acw,\gbm}^{\lasso}$, may be less efficient than $\widehat{\tau}_{\acw}$ due to finite-sample selection error. However, they maintain smaller variance and improved power compared to $\widehat{\tau}_{\aipw}$, regardless of whether the nuisance models are misspecified. When Assumption \ref{assum:exchange_delta} is violated (i.e.,
$\omega=0.3$), $\widehat{\tau}_{\acw}$ becomes biased, leading to
inflated type I error and low power. The Bayesian estimator $\widehat{\tau}_{\ppp}$ requires correct parametric specification of the outcome model and performs poorly when the model omits a key confounder that is imbalanced between data sources. In our simulations, high weights were assigned to the external control subjects, which led to some bias in the treatment effect estimates when $N_{c}$ was small. However, both $\widehat{\tau}_{\acw}^{\lasso}$ and $\widehat{\tau}_{\acw,\gbm}^{\lasso}$ achieve smaller mean squared errors than the trial-only estimator by incorporating external control subjects. In cases where the outcome model is incorrectly specified and $\omega=0.3$, the benefit of using machine learning methods becomes apparent. Specifically, the flexibility of the gradient boosting model ensures the convergence rate assumption for $\widehat{b}_{i}$, i.e., $a_{N}(\widehat{b}_{i}-b_{i,0})=O_{p}(1)$ for a certain sequence $a_{N}$ \citep{zhang2005boosting}. By incorporating compatible external controls more accurately, $\widehat{\tau}_{\acw,\gbm}^{\lasso}$ better controls bias and achieves comparable power levels to $\widehat{\tau}_{\acw}^{\lasso}$, where the adaptive lasso estimation is based on a misspecified linear model that lacks such properties and may not yield power gains. However, the adaptive lasso estimation based on the misspecified linear model lacks such properties and may not provide gains in power. One notable trade-off of our proposed estimators is the slight type I error inflation when $N_{c}$ is small and Assumption \ref{assum:exchange_delta} is violated, which can be attributed to finite-sample selection error and is also observed in \cite{viele2014use}.

\section{Real-data Application \label{sec:application}}

In this section, we present an application of the proposed methodology to investigate the effectiveness of basal Insulin Lispro against regular Insulin Glargine in patients with type I diabetes. When combined with preprandial insulin lispro, basal Insulin Lispro and Insulin Glargine are two long-acting Insulin formulations used for patients with Type I diabetes
mellitus. We analyze the IMAGINE-1 study, a randomized controlled trial where participants were unevenly assigned to either basal Insulin Lispro (treatment group) or Insulin Glargine (control group). Additionally, external control units from the IMAGINE-3 trial were used. In Supplemental Material $\S$\ref{sec:application-II}, we also explore the effectiveness of Solanezumab versus placebo in slowing Alzheimer's Disease progression using external observational data.

Our primary objective is to test the hypothesis of whether basal Insulin Lispro is superior to regular Insulin Glargine at glycemic control for patients with Type I diabetes mellitus. This can be achieved by comparing the deviation of hemoglobin A1c level from baseline after 52 weeks of treatment. Both studies contain a rich set of baseline covariates $X$,
such as age, gender, baseline Hemoglobin A1c (\%), baseline fasting
serum glucose (mmol/L), baseline Triglycerides (mmol/L), baseline
low density lipoprotein cholesterol (mmol/L) and baseline Alanine
Transaminase (U/L). The primary analysis population in IMAGINE-1 were randomized patients who received at least one treatment dose. To mimic the full-analysis population from IMAGINE-1, historical control subjects with missing baseline assessments are discarded from IMAGINE-3. The last observation carried forward is used to impute missing post-baseline outcomes. The IMAGINE-1 study consists of $N_{\rct}=439$ subjects with $286$ in the treated group and $153$ in the control group, while the IMAGINE-3 study includes $N_{\rwd}=444$ patients in the control arm. In our statistical analysis, we first use the baseline covariates $X$ to model the trial inclusion probability
by calibration weighting under the entropy loss function. Next, we
assume a linear heterogeneity treatment effect function for the outcomes
with $X$ as the treatment modifier, and compare the same set of estimators in the simulation study.

Table \ref{tab:est_res-diabetes} reports the estimated results. The
trial-only estimator $\widehat{\tau}_{\aipw}$ shows that the basal Insulin Lispro has a significant treatment effect on reducing the glucose level solely based on the IMAGINE-1 study. Due to potential population bias, the naively integrative estimators $\widehat{\tau}_{\acw}$ and $\widehat{\tau}_{\ppp}$, albeit significant, are slightly different from $\widehat{\tau}_{\aipw}$, which may be subject to possible biases of the external controls. After filtering out the incompatible patients from the external controls by our adaptive lasso selection, the final integrative estimates $\widehat{\tau}_{\acw}^{\lasso}$ and $\widehat{\tau}_{\acw,\gbm}^{\lasso}$ are closer to the benchmark but have narrower confidence intervals. According to our adaptive analysis result, basal Insulin Lispro is significantly more effective than regular Insulin Glargine at glycemic control when used for patients with Type I diabetes mellitus. 

\begin{table}[!ht]
\caption{Point estimates, standard errors, and
95\% confidence intervals of the treatment effect of BIL against regular
GL based on the IMAGINE-1 and IMAGINE-3 studies}
\vspace{0.5em}
\resizebox{\textwidth}{!}{%
\begin{tabular}{lccccc}
\toprule
 & $\widehat{\tau}_{\aipw}$ & $\widehat{\tau}_{\acw}$ & $\widehat{\tau}_{\acw}^{\lasso}$& $\widehat{\tau}_{\acw,\gbm}^{\lasso}$ & $\widehat{\tau}_{\ppp}$\\
 \midrule
Est. (S.E.) & -0.25 (0.072) & -0.22 (0.057) & -0.24 (0.065) & -0.25 (0.070) & -0.27 (0.062)\\
C.I. & (-0.39,-0.11) & (-0.33,-0.11) & (-0.37,-0.08) & (-0.39, -0.12) & (-0.39,-0.15)\\
\bottomrule
\end{tabular}}\label{tab:est_res-diabetes}
\end{table}

\begin{figure}[!h]
    \centering
    \includegraphics[width=.85\linewidth]{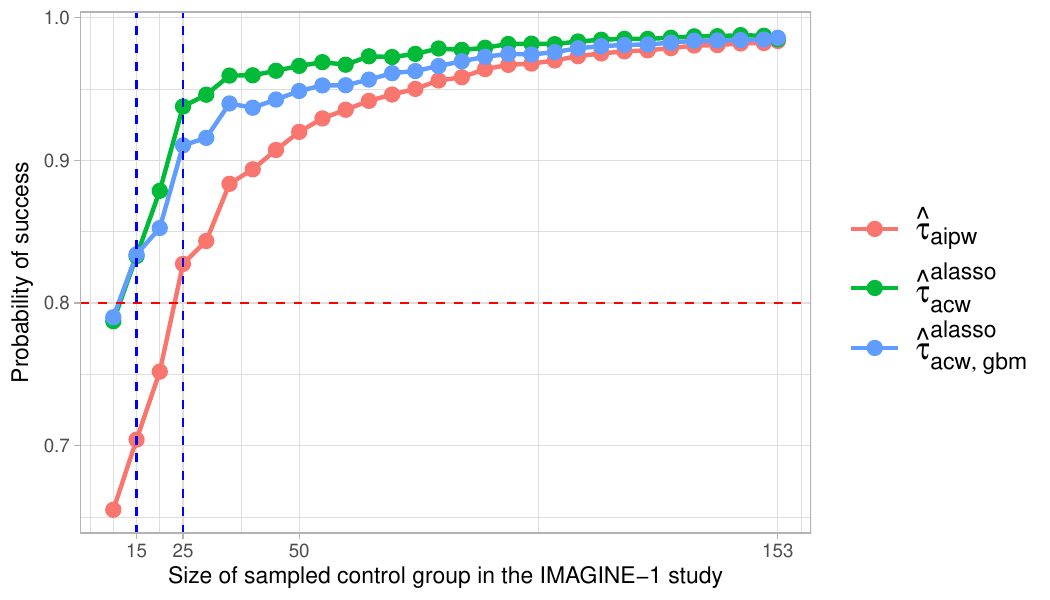}
\caption{\label{fig:AIPW_ACW_figs} Probability of success for detecting $\tau<-0.1$
by $\widehat{\tau}_{\aipw}$ $\widehat{\tau}_{\acw}^{\lasso}$, and $\widehat{\tau}_{\acw,\gbm}^{\lasso}$
with varying control group sizes of the IMAGINE-1 study.}
\end{figure}

Next, we compare the performances of $\widehat{\tau}_{\aipw}$ with our data-adaptive integrative estimates to highlight the advantages
of our dynamic borrowing framework. To accomplish it, we retain the
size of the treatment group but create $100$ sub-samples by randomly
selecting $N_{c}^{s}$ patients from its control group, where $N_{c}^{s}=10,\ldots,153$.
Then, the patients treated with regular Insulin Glargine in the IMAGINE-3 study are augmented to each selected sub-sample and the treatment effect is
evaluated upon the hybrid control arms design. Figure
\ref{fig:AIPW_ACW_figs} presents the average probabilities of successfully
detecting $\tau<-0.1$, the so-called probability of success, against the size of sub-samples. When solely utilizing patients from the IMAGINE-1 study, $\widehat{\tau}_{\aipw}$ produces a probability of success
larger than $0.8$ only if the size of the control group is larger
than $25$. Combined with the IMAGINE-3 study, $\widehat{\tau}_{\acw}^{\lasso}$ and $\widehat{\tau}_{\acw,\gbm}^{\lasso}$ refine
the treatment effect estimation and only $15$ patients are needed in the concurrent control group to attain a probability
of success higher than $0.8$. Therefore, by properly leveraging the
external controls, we may accelerate drug development by decreasing the number of
patients on the concurrent control, thereby reducing the duration
and cost of the clinical trial.

\section{Discussion}\label{sec:discussion}
The interest in the use of external control arms for drug development is becoming more common. However, concerns regarding their quality and validity have limited their use for healthcare decision-making thus far, necessitating careful and appropriate assessment. To adjust for potential selection bias, our proposed method calibrates the covariate moments across two data sources, ensuring that the covariate distributions in both sources match each other. Alternative predictive model-based strategies are applicable when only a subset of covariates is shared \citep{stuart2011use, tipton2014generalizable}. To address differences in outcomes, we select comparable external subsets based on the adaptive lasso penalty. Alternative penalties can be considered if the selection consistency property is attained, such as the smoothly clipped absolute deviation penalty \citep{fan2001variable}. Moreover, our framework can be easily extended to augment observational studies with external data, which may require additional modeling and assumptions to achieve double robustness. Slight type I error inflation is observed in our simulations when the concurrent control group is small, attributed to selection error in finite samples. One future direction will be to rigorously construct a data-adaptive confidence interval to account for finite-sample selection uncertainty without being overly conservative \citep{lee2016exact, tibshirani2016exact}. Other future directions include extending the proposed integrated inferential framework to survival outcomes \citep{lee2022doubly}, estimating heterogeneous treatment effects \citep{wu2022integrative, yang2022elastic}, and combining probability and non-probability samples \citep{yang2020doubly, gao2023pretest}.

\section*{Acknowledgement}
This project is supported by the Food and Drug Administration (FDA) of the U.S. Department of Health and Human Services (HHS) as part of a financial assistance award U01FD007934 totaling $\$1,674,013$ over two years funded by FDA/HHS. It is also supported by the National Institute On Aging of the National Institutes of Health under Award Number R01AG06688, totaling $\$1,565,763$ over four years. The contents are those of the authors and do not necessarily represent the official views of, nor an endorsement by, FDA/HHS, the National Institutes of Health, or the U.S. Government.

\section*{Supplementary material}
Supplementary material available at \Bka\ online includes all technical proofs, additional simulation results, and other real-data applications. An open-source software R package is available for implementing our proposed methodology at 
 \url{https://github.com/IntegrativeStats/SelectiveIntegrative}.

\bibliographystyle{dcu}
\bibliography{ci}

\newpage
\appendix

\section*{Supplementary Material}
We include all the technical details in this section. In specific, Section \ref{sec:proof_summary} provides the proofs of Theorems \ref{thm:tau_EIF},
\ref{thm:rate_robustness}, \ref{thm:acw-final-point}, \ref{thm:acw-final-CI}, extensions to multiple external controls, and other technical Lemmas. Section \ref{sec:More-Simulation-Results} presents additional simulation
results related to other bias-generating concerns raised by FDA guidance. Section \ref{sec:application-II} presents an additional real-data application on the progression of mild Alzheimer’s Disease.

\section{Technical Proofs and Details}\label{sec:proof_summary}

\subsection{Proof of Theorem \ref{thm:tau_EIF}\label{subsec:proof-thm-EIF}}
To identify the parameter $\mu_{0}$ with the observed data $\{V_{i}:i\in\mathcal{R}\cup\mathcal{E}\}$,
we have
\begin{align}
\mu_{0} & =\E[\E\{Y(0)\mid X,R=1\}\mid R=1]\nonumber \\
 & =\E[\E\{Y(0)\mid X,R=0\}\mid R=1]\nonumber \\
 & =\E[\E(Y\mid X,R=0)\mid R=1]\nonumber \\
 & =\frac{1}{P(R=1)}\E\left\{R \E(Y\mid X, R=0)\right\} \nonumber \\
 & =\frac{1}{P(R=1)}\E\left\{P(R=1\mid X) \mu_0(X)\right\},\label{eq:identified}
\end{align}
where $\mu_0(X) = \E(Y\mid X, R=1, A= 0) = \E(Y\mid X, R=0)$ under Assumption \ref{assum:exchange_delta}. Let $V=(X,A,Y,R)$ be a vector of random variables, the full observed data distribution is
\begin{align*}
f(V) & =f(X)P(R=1\mid X)^{R}P(R=0\mid X)^{1-R}\\
 & \times P(A=1\mid X,R=1)^{RA}P(A=0\mid X,R=1)^{R(1-A)}\\
 & \times f(Y\mid X,R=1, A=1)^{RA}f(Y\mid X,R=1, A=0)^{R(1-A)}\\
 &\times f(Y\mid X,R=0)^{1-R}.
\end{align*}
To derive the efficient influence function, we resort to the method
of parametric submodel in \citet{bickel1993efficient}. Let $\{f_{t}(V):t\in\mathbb{R}\}$
be a regular parametric submodel and the truth is evaluated at $t=0$,
i.e., $f_{t}(V)\mid_{t=0}=f(V)$. Therefore, the observed score function based
on single observed $V$ under Assumptions \ref{a:ign} and \ref{assum:exchange_delta}
is derived by the pathwise derivatives of $\log f_{t}(V)$ with respect
to $t$: 
\begin{align*}
S_{t}(V) & =S_{t}(X)+\frac{R-P(R=1\mid X)}{P(R=1\mid X)\{1-P(R=1\mid X)\}}\frac{\partial P_{t}(R=1\mid X)}{\partial t}\\
 & +\frac{R\{A-\pi_{A}(X)\}}{\pi_{A}(X)\{1-\pi_{A}(X)\}}\frac{\partial P_t(A=1\mid X, R=1)}{\partial t}\\
 &+RAS_{t}(Y\mid X,A=1, R=1)
 +R(1-A)S_{t}(Y\mid X,R=1,A=0)\\
 & +(1-R)S_{t}(Y\mid X,R=0),
\end{align*}
where $S_{t}(X)=\partial\log f_{t}(X)/\partial t$, $S_{t}(Y\mid X,R=1, A=a)=\partial\log f(Y\mid X,R=1, A=a)/\partial t$
for $a=0,1$, and $S_{t}(Y\mid X,R=0)=\partial\log f(Y\mid X,R=0)/\partial t$.
By the observed score function, the tangent space $\mathcal{T}$ can be
constructed by $\mathcal{T} =\mathcal{T}_{1}+\mathcal{T}_{2}+\mathcal{T}_{3}+\mathcal{T}_{4}$, where 
\begin{align*}
\mathcal{T}_{1} & =\{\Gamma(X):\int\Gamma(x)f(x)dx=0\},\\
\mathcal{T}_{2} & =\left\{ \{R-P(R=1\mid X)\}a(X)\right\} ,\quad\mathcal{T}_{3}=\left\{ R(A-\pi_{A})b(X)\right\} ,\\
\mathcal{T}_{4} & = \mathcal{T}_{41}\cap \mathcal{T}_{42} =\left\{ \Gamma(Y, X, R, A):\int\Gamma(y, X, R, A)f(y\mid X, R, A)dy=0\right\} \\
&\cap \left\{ \Gamma(Y, X, R, A):
\E\left[\left\{
\frac{(1-R)Y}{P(R=0\mid X)}- 
\frac{R(1-A)Y}{P(R=1, A=0\mid X)}\right\}\Gamma(Y, X, R, A)\mid X\right]=0\right\},
\end{align*}
for any two arbitrary square-integrable measurable functions $a(X),b(X)$. The tangent space $\mathcal{T}_{42}$ is induced by the restricted moment model due to Assumption \ref{assum:exchange_delta}, where $\E(Y\mid X, R=1, A= 0) = \E(Y\mid X, R=0)$. 

From the semiparametric theory, the efficient influence function $\psi_{\mu_0,\eff}(V)$ for $\mu_0$ must satisfy $\partial \E\{\mu_{0,t}(X)\}/\partial t \big|_{t=0}= \E\{\psi_{\mu_0,\eff}(V) S(V)\}$ and belongs to the tangent space $\mathcal{T}$. Based on the formula in (\ref{eq:identified}), we have $\mu_{0} = \E\left\{ P(R=1\mid X)\mu_{0}(X)\right\}/P(R=1)$, which is in a ratio form with numerator $N=\E\left\{ P(R=1\mid X)\mu_{0}(X)\right\} $
and denominator $D=P(R=1)$. Therefore, we will first define the efficient influence functions of the numerator and denominator,
and then combine them to have the final efficient influence function for $\mu_{0}$.

Let $N_{t}$
and $D_{t}$ denote $N$ and $D$ being evaluated at the submodel
$f_{t}(V)$. For the numerator $N_{t}$, the semiparametric efficient
influence function is the pathwise derivative of the parameter of interest:
\begin{eqnarray*}
\frac{\partial N_{t}}{\partial t}\mid_{t=0} & = & \E\left\{ P(R=1\mid X)\mu_{0}(X)S(X)\right\} \\
 &  & +\E_{t}\left[\frac{\partial P_{t}(R=1\mid X)}{\partial t}\mu_{0,t}(X)\right]\bigg|_{t=0}+\E_{t}\left[\frac{\partial\mu_{0,t}(X)}{\partial t}P(R=1\mid X)\right]\bigg|_{t=0}.
\end{eqnarray*}
Next, we show that the pathwise derivative in the second part is
\begin{align}
\frac{\partial P_{t}(R=1\mid X)}{\partial t}\bigg|_{t=0} & =\frac{\partial}{\partial t}\E_{t}(R\mid X)\bigg|_{t=0}\nonumber \\
 & =\E\left[\{R-P(R=1\mid X)\}S(A,Y,R\mid X)\mid X\right].\label{eq:partial_P}
\end{align}
But the pathwise derivative $\partial\mu_{0,t}(X)/\partial t$ in the third part can be derived in two different ways under the conditional mean exchangeability assumption by (\ref{eq:pathwise_mu0_way1}) and (\ref{eq:pathwise_mu0_way2}):
\begin{align}
\frac{\partial\mu_{0,t}(X)}{\partial t}\bigg|_{t=0} & =\left\{ \int y\frac{\partial}{\partial t}f_{t}(y\mid X,R=0)dy\right\} \bigg|_{t=0}\nonumber\\
 & = \left\{ \int yS(y\mid X,R=0)f_{t}(y\mid X,R=0)dy\right\} \bigg|_{t=0}\nonumber\\
 & =  \E\left\{ YS(Y\mid X,R=0)\mid X,R=0\right\} \nonumber\\
 & = \E\left[\left\{ Y-\mu_{0}(X)\right\} S(Y\mid X,R=0)\mid X,R=0\right]\nonumber\\
 & = \E\left[\frac{(1-R)\left\{ Y-\mu_{0}(X)\right\} S(Y\mid X,R)}{P(R=0\mid X)}\bigg|X\right]\label{eq:pathwise_mu0_way1},
\end{align}
and
\begin{align}
\frac{\partial\mu_{0,t}(X)}{\partial t}\bigg|_{t=0} & = \E\left[\left\{ Y-\mu_{0}(X)\right\} S(Y\mid X,R=0)\mid X,R=0\right]\nonumber\\
 & =\E\left[\left\{ Y-\mu_{0}(X)\right\} S(Y\mid X,R=1,A=0)\mid X,R=1,A=0\right]\nonumber\\
 & =\E\left[\frac{R(1-A)\left\{ Y-\mu_{0}(X)\right\} S(Y\mid X,R,A)}{P(R=1,A=0\mid X)}\bigg|X\right]\label{eq:pathwise_mu0_way2}.
\end{align}
By combining the formulas (\ref{eq:pathwise_mu0_way1}) and (\ref{eq:pathwise_mu0_way2}), the pathwise derivative of $\mu_{0,t}(X)$ in general should be
\begin{equation}
    \begin{split}
        \frac{\partial\mu_{0,t}(X)}{\partial t}\bigg|_{t=0} &
 = \E\left[\frac{C_1}{C_1 + C_2}\frac{(1-R)\left\{ Y-\mu_{0}(X)\right\} S(Y\mid X,R)}{P(R=0\mid X)}\bigg|X\right] \\
 & + \E\left[\frac{C_2}{C_1 + C_2}\frac{R(1-A)\left\{ Y-\mu_{0}(X)\right\} S(Y\mid X,R,A)}{P(R=1,A=0\mid X)}\bigg|X\right],
    \end{split}
    \label{eq:pathwise_mu0_general}
\end{equation}
where $C_1=C_1(X)$ and $C_2 = C_2(X)$ are two arbitrary functions of $X$. To obtain the efficient influence function of $N$, we need to find the proper $C_1$ and $C_2$ such that the third term belongs to the tangent space $\mathcal{T}_4$, which satisfies:
\begin{align*}
    \E&\left(\left[{{C_1} (1-R)q(X)\left\{ Y-\mu_{0}(X)\right\}}
    +{{C_2} R(1-A)\frac{Y-\mu_{0}(X)}{1-\pi_A(X)}} 
\right]\right.
\\
&\left.\times \left\{\frac{(1-R)Y}{P(R=0\mid X)}- 
\frac{R(1-A)Y}{P(R=1, A=0\mid X)}\right\}
\mid X
\right) = 0.
\end{align*}
By algebra, we can show that 
\begin{align}
    \frac{C_1}{C_2} &= 
\frac{\E \left[R^2(1-A)^2\{Y-\mu_0(X)\}Y\mid X\right]}{\{1-\pi_A(X)\}P(R=1,A=0\mid X)}
\bigg/
\frac{\E \left[(1-R)^2q(X)\{Y-\mu_0(X)\}Y\mid X\right]}{P(R=0\mid X)}\nonumber\\
&= \frac{r(X)}{\{1-\pi_A(X)\}q(X)},\label{eq:C1_over_C2}
\end{align}
where $q(X)=P(R=1\mid X)/P(R=0\mid X)$ and $r(X)=\text{var}(Y\mid X,R=1,A=0)/\text{var}(Y\mid X,R=0)$.
Plugging (\ref{eq:pathwise_mu0_general}) and (\ref{eq:C1_over_C2}) back, we can show that the desired pathwise derivative of $\mu_{0,t}(X)$ is
$$
\frac{(1-R)r(X)q(X)\left\{ Y-\mu_{0}(X)\right\}}{r(X)+\{1-\pi_{A}(X)\}q(X)} +\frac{R(1-A)q(X)\left\{ Y-\mu_{0}(X)\right\}}{r(X)+\{1-\pi_{A}(X)\}q(X)} \in \mathcal{T}_4.
$$
Thus, we verify that each part of $\partial N_t/\partial t\mid_{t=0}$ belongs to the tangent space $\mathcal{T}$ and the efficient influence function for $N$ is
\begin{eqnarray*}
\psi_{N} & = & P(R=1\mid X)\mu_{0}(X)-N+\left\{ R-P(R=1\mid X)\right\} \mu_{0}(X)\\
 &  &+\frac{(1-R)r(X)q(X)\left\{ Y-\mu_{0}(X)\right\}}{r(X)+\{1-\pi_{A}(X)\}q(X)} +\frac{R(1-A)q(X)\left\{ Y-\mu_{0}(X)\right\}}{r(X)+\{1-\pi_{A}(X)\}q(X)}  \\
 & = & \phi_{N}-N.
\end{eqnarray*}
For the denominator $D_{t}$,
we have $\partial D_{t}/\partial t\mid_{t=0}=\E\{R\partial P_{t}(R=1)/\partial t\}|_{t=0}$
and thus, the semiparametric efficient influence function based on Lemma S2 in
\citet{jiang2020multiply} is 
\begin{align*}
\psi_{\mu_{0},\eff}(V) & =\frac{\phi_{N}-\mu_{0}R}{P(R=1)}\\
 & =
 \underbrace{\frac{P(R=1\mid X)\{\mu_0(X)-\mu_0\}}{P(R=1)}}_{\in \mathcal{T}_1} + 
 \underbrace{\frac{\{R - P(R=1\mid X)\}\{\mu_0(X)-\mu_0\}}{P(R=1)}}_{\in \mathcal{T}_2}\\
 & + \underbrace{\frac{1-R}{P(R=1)}\frac{q(X)r(X)\left\{ Y-\mu_{0}(X)\right\}}{r(X)+\{1-\pi_{A}(X)\}q(X)} +\frac{R(1-A)}{P(R=1)}\frac{q(X)\left\{ Y-\mu_{0}(X)\right\}}{r(X)+\{1-\pi_{A}(X)\}q(X)}}_{\in \mathcal{T}_4},
\end{align*}
which has the semiparametric efficiency bound as 
\begin{align*}
 & V_{\mu_{0},\eff}=\left[\E\left\{ \frac{\partial\psi_{\mu_{0},\eff}(V)}{\partial\mu_{0}}\right\} \right]^{-2}\E\{\psi_{\mu_{0},\eff}^{2}(V)\}=\E\{\psi_{\mu_{0},\eff}^{2}(V)\},
\end{align*}
and $\E\left\{ \partial\psi_{\mu_{0},\eff}(V)/\partial\mu_{0}\right\} =-\E\left\{ R/P(R=1)\right\} =-1$. Following the similar argument in Section \ref{subsec:proof-thm-EIF},
the semiparametric efficient influence function $\psi_{\tau,\eff}(X,A,Y,R;\tau)$ for $\tau$ can be derived 
\begin{align*}
\psi_{\tau,\eff}(V) & =
\underbrace{\frac{P(R=1\mid X)\{\mu_1(X) - \mu_0(X) - \tau\}}{P(R=1)}}_{\in \mathcal{T}_1}\\
& + 
\underbrace{\frac{\{R - P(R=1\mid X)\}\{\mu_1(X) - \mu_0(X) - \tau\}}{P(R=1)}}_{\in \mathcal{T}_2}+ 
\underbrace{\frac{RA\{Y-\mu_{1}(X)\}}{\pi_A(X)}}_{\in \mathcal{T}_4}
\\& 
\underbrace{-\frac{1-R}{P(R=1)}\frac{q(X)r(X)\{Y-\mu_{0}(X)\}}{r(X)+\{1-\pi_{A}(X)\}q(X)}-\frac{R(1-A)}{P(R=1)}\frac{q(X)\left\{ Y-\mu_{0}(X)\right\} }{r(X)+\{1-\pi_{A}(X)\}q(X)}}_{\in \mathcal{T}_4}.
\end{align*}
By setting the empirical efficient influence function under the observed data $(X,A,Y,R)$ equal to zero, the semiparametric efficient estimator of $\tau$ is obtained by 
\begin{align*}
\widehat{\tau} & =N_{\mathcal{R}}^{-1}\sum_{i\in\mathcal{R}\cup\mathcal{E}}R_{i}\left[\frac{A_{i}\{Y_{i}-\widehat{\mu}_{1}(X_{i})\}}{\pi_{A}(X_{i})}+\widehat{\mu}_{1}(X_{i})-\widehat{\mu}_{0}(X_{i})\right]\\
 & -N_{\mathcal{R}}^{-1}\sum_{i\in\mathcal{R}\cup\mathcal{E}}\frac{(1-R_{i})\widehat{q}(X_{i})\widehat{r}(X_{i})\left\{ Y_{i}-\widehat{\mu}_{0}(X_{i})\right\}}{\widehat{r}(X_{i})+\{1-\pi_{A}(X_{i})\}\widehat{q}(X_{i})} -N_{\mathcal{R}}^{-1}\sum_{i\in\mathcal{R}\cup\mathcal{E}}\frac{R_{i}(1-A_{i})\widehat{q}(X_{i})\left\{ Y_{i}-\widehat{\mu}_{0}(X_{i})\right\}}{\widehat{r}(X_{i})+\{1-\pi_{A}(X_{i})\}\widehat{q}(X_{i})},
\end{align*}
which achieves the semiparametric efficiency bound $\bV_{\tau,\eff}$ if all the nuisance functions are correctly specified as illustrated in Theorem \ref{thm:rate_robustness}.

\subsection{Proof of Double Robustness of $\widehat{\tau}_{\acw}$ \label{subsec:proof_dr_acw}}
To prove the double robustness of $\widehat{\tau}_{\acw}$, we aim
to show that $\E\{\psi_{\tau,\eff}(V)\}=0$ if either $\mu_{0}(X)$
or $q(X)$ is correctly specified regardless of whether $r(X)$ is correct or not.

When $\mu_{0}(X)$ is correctly specified but $q(X)$ is incorrectly specified: Under this scenario, we have $\widehat{\mu}_{0}(X)\rightarrow\mu_{0}(X)$
and $\widehat{q}(X)\rightarrow q^{w}(X)$, where $q^{w}(X)$ is an
arbitrary function. Then, we can show that 
\begin{align}
 & P(R=1)\E\{\psi_{\tau,\eff}(V)\}\nonumber\\
 & =
 \E\left[
 R\left\{
 \frac{AY}{\pi_A(X)} - \mu_1
 \right\}
 +
 R\widehat{\mu}_1(X)\left\{
 1-\frac{A}{\pi_A(X)}
 \right\}
 \right]\label{eq:mu0_part1}\\
 & -\E[R\{\widehat{\mu}_{0}(X)-\mu_{0}\}]\label{eq:mu0_part2}\\
 & -\E\left[\frac{(1-R)\widehat{q}(X)r(X)\{Y-\widehat{\mu}_0(X)\}}{r(X)+\{1-\pi_{A}(X_{i})\}\widehat{q}(X)}\right] -\E\left[\frac{R(1-A)\widehat{q}(X)\left\{ Y-\widehat{\mu}_0(X)\right\}}{r(X)+\{1-\pi_{A}(X_{i})\}\widehat{q}(X)} \right],\label{eq:mu0_part3}
\end{align}
where (\ref{eq:mu0_part1}) is always consistent with zero as the true $\pi_A(X)$ is known for the randomized clinical trial. The second part (\ref{eq:mu0_part2}) is consistent with 
\begin{align*}
\E[R\{\mu_{0}(X)-\mu_{0}\}] & =\E\{\mu_{0}(X)-\mu_{0}\mid R=1\}P(R=1)=0
\end{align*}
 as $\mu_{0}=\E(Y\mid R=1, A=0)=\E\{\mu_{0}(X)\mid R=1\}$. For the rest parts (\ref{eq:mu0_part3}), they are consistent with zero under the Assumption \ref{assum:exchange_delta}:
 \begin{align*}
 & \E\left[\frac{(1-R)q^{w}(X)r(X)\{Y-\mu_{0}(X)\}}{r(X)+\{1-\pi_{A}(X)\}q^{w}(X)}\right]\\
= & \E\left[\frac{P(R=0\mid X)q^{w}(X)r(X)}{r(X)+\{1-\pi_{A}(X)\}q^{w}(X)}\left\{ \E(Y\mid X,R=0)-\mu_{0}(X)\right\} \right]\\
= & \E\left[\frac{P(R=0\mid X)q^{w}(X)r(X)}{r(X)+\{1-\pi_{A}(X)\}q^{w}(X)}\{\E(Y\mid X,R=1,A=0)-\mu_{0}(X)\}\right]=0,
\end{align*}
and
\begin{align*}
 & \E\left[\frac{R(1-A)q^{w}(X)\left\{ Y-\mu_{0}(X)\right\}}{r(X)+\{1-\pi_{A}(X)\}q^{w}(X)} \right]\\
= & \E\left(\frac{q^{w}(X)P(R=1,A=0\mid X)}{r(X)+\{1-\pi_{A}(X)\}q^{w}(X)}\left[\E(Y\mid X,R=1)-\mu_{0}(X)\right]\right)=0.
\end{align*}
Next, when $q(X)$ is correctly specified but $\mu(X)$ is incorrectly specified: Under this scenario, we have $\widehat{\mu}_{0}(X)\rightarrow\mu_{0}^{w}(X)$
and $\widehat{q}(X)\rightarrow q(X)$, where $\mu_{0}^{w}(X)$ is
an arbitrary function. Then, (\ref{eq:mu0_part3}) is consistent with
\begin{align*}
 & \E\left[\frac{(1-R)q(X)r(X)\{Y-\mu_{0}^{w}(X)\}}{r(X)+\{1-\pi_{A}(X)\}q(X)}+\frac{R(1-A)q(X)\{Y-\mu_{0}^{w}(X)\}}{r(X)+\{1-\pi_{A}(X)\}q(X)}\right]\\
= & \E\left[\frac{P(R=1\mid X)r(X)}{r(X)+\{1-\pi_{A}(X)\}q(X)}\{\E(Y\mid X,R=0)-\mu_{0}^{w}(X)\}\right]\\
 & +\E\left[\frac{P(R=1\mid X)\{1-\pi_{A}(X)\}q(X)}{r(X)+\{1-\pi_{A}(X)\}q(X)}\{\E(Y\mid X,R=1,A=0\}-\mu_{0}^{w}(X)\}\right]\\
= & \E\left[P(R=1\mid X)\{\E(Y\mid X,R=1, A=0)-\mu_{0}^{w}(X)\}\right]\\
= & \E\left[R\{\mu_{0}(X)-\mu_{0}^{w}(X)\}\right],
\end{align*}
which is canceled out with (\ref{eq:mu0_part2}). Thus, $\E\{\psi_{\tau,\eff}(V)\}$ is an unbiased estimating
equation for $\tau$ if $q(X)$ is correctly specified. 

When all the nuisance functions are correctly specified, we show in Theorem \ref{thm:rate_robustness}
that the influence function of $\widehat{\tau}_{\acw}$ is the same
as the efficient influence function in Theorem \ref{thm:tau_EIF}. Thus, it attains the semiparametric
efficiency bound $\bV_{\tau,\eff}$. To derive a consistent variance
estimator for $\widehat{\tau}_{\acw}$ under the parametric modeling,
we take the Taylor's series of $\widehat{\tau}_{\acw}-\tau$ as
\begin{align*}
\widehat{\tau}_{\acw}-\tau & =N^{-1}\sum_{i\in\mathcal{R}\cup\mathcal{E}}\psi_{\tau,\acw}(V_{i};\widehat{\mu}_{1},\widehat{\mu}_{0},\widehat{q},\widehat{r})\\
 & =N^{-1}\sum_{i\in\mathcal{R}\cup\mathcal{E}}\psi_{\tau,\acw}^{\text{dr}}(V_{i};\widehat{\mu}_{1},\widehat{\mu}_{0},\widehat{q},\widehat{r})+o_{p}(N^{-1/2}),
\end{align*}
where
\begin{align*}
 & \psi_{\tau,\acw}(V_{i};\widehat{\mu}_{1},\widehat{\mu}_{0},\widehat{q},\widehat{r}) =\frac{R}{P(R=1)}\left[\{\widehat{\mu}_{1}(X) -\widehat{\mu}_{0}(X) - \tau\}+\frac{A\{Y-\widehat{\mu}_{1}(X)\}}{\pi_A(X)}\right]\\
 & -\frac{1-R}{P(R=1)}\frac{\widehat{q}(X)\widehat{r}(X)\{Y-\widehat{\mu}_{0}(X)\}}{\widehat{r}(X)+\{1-\pi_{A}(X)\}\widehat{q}(X)} -\frac{R(1-A)}{P(R=1)}\frac{\widehat{q}(X)\left\{Y-\widehat{\mu}_{0}(X)\right\}}{\widehat{r}(X)+\{1-\pi_{A}(X)\}\widehat{q}(X)},
\end{align*}
and $\psi_{\tau,\acw}^{\text{dr}}(V_{i};\widehat{\mu}_{1},\widehat{\mu}_{0},\widehat{q},\widehat{r})$
motivates the variance estimation presented in Remark \ref{rmk:double-robust-variance}.

\begin{remark}

\label{rmk:double-robust-variance} Under Assumptions \ref{a:ign},
\ref{assum:exchange_delta} and \ref{assum:rate-conditions}, the
asymptotic variance $\bV_{\tau}$ for $N^{1/2}(\widehat{\tau}_{\acw}-\tau)$
can be consistently estimated by $\widehat{\bV}_{\tau}$, where $\widehat{\bV}_{\tau}=N^{-1}\sum_{i\in\mathcal{R}\cup\mathcal{E}}\left\{ \psi_{\tau,\acw}^{\text{dr}}(V_{i};\widehat{\mu}_{1},\widehat{\mu}_{0},\widehat{q},\widehat{r})\right\} ^{2}$. If $\|\widehat{r}(X)-r(X)\|=o_p(1)$, $\widehat{\bV}_{\tau}$ converges to the semiparametric efficient bound $\bV_{\tau,\eff}$.
\end{remark}
\begin{proof}[of of Remark \ref{rmk:double-robust-variance}]

Let the nuisance functions $(\widehat{\mu}_{1},\widehat{\mu}_{0},\widehat{q})$
be characterized by the parameter $\theta=(\beta_{0},\beta_{1},\eta)$, solving by the following estimation equations:
\begin{align*}
S_{\beta_{0}}(V;\beta_{0}) & =(1-A)R\{Y-\mu_{0}(X;\beta_{0})\}\frac{\partial\mu_{0}(X;\beta_{0})}{\partial\beta_{0}},\\
S_{\beta_{1}}(V;\beta_{1})&=RA\{Y-\mu_{1}(X;\beta_{1})\}\frac{\partial\mu_{1}(X;\beta_{1})}{\partial\beta_{1}},\\
S_{\eta}(V;\eta) & =\{(1-R)q(X;\eta)-R\}X.
\end{align*}
Based on standard Taylor first-order expansion, $\widehat{\tau}_{\acw}$
has the influence function
\begin{align*}
 & \psi_{\tau,\acw}^{\text{dr}}(V;\theta)=\psi_{\tau,\acw}(V;\theta)\\
 & -\E\left\{ \frac{\partial\psi_{\tau,\acw}(V;\theta)}{\partial\beta_{1}^{\T}}\right\} \left\{ \E\frac{\partial S_{\beta_{1}}(X,A,Y;\beta_{1})}{\partial\beta_{1}}\right\} ^{-1}S_{\beta_{1}}(V;\beta_{1})\\
 & -\E\left\{ \frac{\partial\psi_{\tau,\acw}(V;\theta)}{\partial\beta_{0}^{\T}}\right\} \left\{ \E\frac{\partial S_{\beta_{0}}(X,A,Y;\beta_{0})}{\partial\beta_{0}}\right\} ^{-1}S_{\beta_{0}}(V;\beta_{0})\\
 & -\E\left\{ \frac{\partial\psi_{\tau,\acw}(V;\theta)}{\partial\eta^{\T}}\right\} \left\{ \E\frac{\partial S_{\eta}(X,A,Y;\eta)}{\partial\eta}\right\} ^{-1}S_{\eta}(V;\eta),
\end{align*}
where $\psi_{\tau,\acw}^{\text{dr}}(V;\theta)$ is considered by adding
the derivative of $\psi_{\tau,\acw}(V;\theta)$ in the direction of
$\widehat{\theta}-\theta$ to the original influence function
$\psi_{\tau,\acw}(V;\theta)$. Consequently, a doubly robust variance
estimator for $\bV_{\tau,\acw}$ can be derived by 
\[
\widehat{\bV}_{\tau}=N^{-1}\sum_{i\in\mathcal{R}\cup\mathcal{E}}\left\{ \psi_{\tau,\acw}^{\text{dr}}(V_{i};\widehat{\theta})\right\}^{2}.
\]
Under Assumption \ref{assum:rate-conditions}, where $\widehat{\mu}_{a}(X),a=0,1$
and $\widehat{q}(X)$ are characterized by flexible modeling, the bias of the semiparametric efficient influence function is equal to the product
of the estimation errors of two nuisance functions, which vanishes
asymptotically under Assumption \ref{assum:rate-conditions}. Therefore,
the estimated efficient influence function $\psi_{\tau,\acw}(V;\widehat{\mu}_{1},\widehat{\mu}_{0},\widehat{q})$
can be directly used for the variance estimation. 
\end{proof}
Another alternative variance estimation is the bootstrap strategy
based on the original observations $\{V_{i}:i\in\mathcal{R}\cup\mathcal{E}\}$
or the asymptotic linear terms $\psi_{\tau,\acw}^{\text{dr}}(V_{i};\widehat{\mu}_{1},\widehat{\mu}_{0},\widehat{q},\widehat{r})$;
see \citet{otsu2017bootstrap} for more details. 

\subsection{Proof of Theorem \ref{thm:rate_robustness} \label{subsec:Proof-of-Theorem-rate_robust}}
\begin{assumption}\label{assum:regularity-eff}

Let $V_{i}=(X_{i},A_{i},Y_{i},R_{i})$, the following regularity conditions
hold:
\begin{enumerate}
\item[a)]  $\psi_{\tau,\eff}(V)$ belongs to a Donsker class \citep{van2000asymptotic}.
\item[b)]  $\psi_{\tau,\eff}(V)$ is differentiable in $\tau$ and $\E\{\partial\psi_{\tau,\eff}(V)/\partial\tau^{\T}\}$
exists and is invertible.
\end{enumerate}
\end{assumption}

Assumption \ref{assum:regularity-eff} regularizes the complexity
of the functional space. Noted that Assumption \ref{assum:regularity-eff} a) may not be required if a cross-fitting procedure is used as an alternative. We illustrate Theorem \ref{thm:rate_robustness}
by choosing the flexible data-adaptive modeling to be the method of
sieves \citep{chen2007large}. Based on the covariate $X\in\mathbb{R}^{d_{X}}$,
we consider a $d$-vector basis function $g(X)=\{g_{1}(X),\cdots,g_{d}(X)\}^{\T}$
and approximate $\{q(X),\mu(X)\}$ by the generalized sieves functions
\[
q(X;\eta^{*})=h_{q}\{\eta^{*\T}g(X)\},\quad\mu_{a}(X;\beta_{a}^{*})=h_{\mu,a}\{\beta_{a}^{*\T}g(X)\},
\]
where $h_{q}(\cdot)$ is the link function for calibration weighting
induced by the objective function, e.g., for the entropy balancing,
$h_{q}(\cdot)$ is an experiential function, and for the maximum entropy,
$h_{q}(\cdot)$ is an expit function, $h_{\mu,a}(\cdot)$ is the link
function for $\mu_{a}$, and 
\begin{equation}
\eta^{*}=\arg\min_{\eta}\E[q(X)-h_{q}\{\eta^{\T}g(X)\}]^{2},\quad\beta_{a}^{*}=\arg\min_{\beta_{a}}\E[\mu_{a}(X)-h_{\mu,a}\{\beta_{a}^{\T}g(X)\}]^{2}.\label{eq:obj_eta_beta}
\end{equation}
Next, we present the regularity conditions for the sieves estimator,
under which Theorem \ref{thm:rate_robustness} holds.

\begin{assumption} \label{assum:regularity-seives}

The following regularity conditions hold:
\begin{enumerate}
\item[a)]  (Distribution of $X$) Let $\mathcal{X}$ be the support of $X$
and is a Cartesian product of compact intervals. The density of $X$,
$f(X)$, is bounded above and below away from $0$ on $\mathcal{X}$.
\item[b)]  (Bounded moment) The second moment of the potential outcomes are
finite, i.e., $\E\{Y(a)^{2}\}<\infty,$ for $a=0,1$.
\item[c)]  (Functional smoothness) $q(X)$ is $s_{1}$-times continuously differentiable,
and $\mu_{a}(X)$ is $s_{2}$-times continuously differentiable $\forall X\in\mathcal{X}$;
let $s_{0}=\min(s_{1},s_{2})$ and $s>3d_{X}$.
\item[d)]  (Basis functions) There exist constant $l$ and $u$ such that 
\[
l\leq\rho_{\min}\{g(X)^{\T}g(X)\}\leq\rho_{\max}\{g(X)^{\T}g(X)\}\leq u,
\]
almost surely where $\rho_{\min}(\cdot)$ and $\rho_{\max}(\cdot)$
denote the minimum and maximum eigenvalues of a matrix. The number
of basis function $d$ satisfies $d=O(N^{\nu})$, where $d_{X}/(s_{0}-d_{X})<\nu<1/4$.
\end{enumerate}
\end{assumption}
Under the regularity conditions in Assumption \ref{assum:regularity-seives}
and following \citet{newey1997convergence}, the bounds for the bias
between the true functional and the sieves approximation are 
\begin{align*}
 & \sup_{X\in\mathcal{X}}|q(X)-h_{q}\{\eta^{\T}g(X)\}|=O\{d^{1-s_{1}/(2d_{X})}\},\\
 & \sup_{X\in\mathcal{X}}|\mu_{a}(X)-h_{\mu,a}\{\beta_{a}^{\T}g(X)\}|=O\{d^{1-s_{2}/(2d_{X})}\},
\end{align*}
and $O\{d^{1-s_{1}/(2d_{X})}\}=o(N^{-1/4})$ under Assumption \ref{assum:regularity-seives}d).
Besides, since $d^{4}=o(N)$, the variances of the sieves approximations
are $O(d/N)=o(N^{-1/2})$. Then under Assumption \ref{assum:regularity-seives}d),
we have i) $\|\widehat{\mu}_{0}(X)-\mu_{0}(X)\|=o_{p}(1)$ and $\|\widehat{q}(X)-q(X)\|=o_{p}(1)$;
(ii) $\|\widehat{\mu}_{0}(X)-\mu_{0}(X)\|\times\|\widehat{q}(X)-q(X)\|=o_{p}(N^{-1/2})$. 

Following the empirical process literature, we denote $\widehat{\pr}$
as the empirical measure over the combined trial data and external controls, i.e.,
$\widehat{\pr}\{h(V)\}=N^{-1}\sum_{i\in\mathcal{R}\cup\mathcal{E}}h(V_{i})$.
Also, we let $\pr$ denotes the expectation over the data generative
distribution, i.e., $\pr\{h(V)\}=\int h(V)d\pr$. Under Assumption
\ref{assum:regularity-eff}, by the standard Taylor expansion, we
have 
\begin{align*}
\widehat{\tau}_{\acw}-\tau_{0} & =-\left[\E\left\{ \frac{\partial\psi_{\tau,\acw}(V;\mu_{1},\mu_{0},q,r)}{\partial\tau}\right\} \right]^{-1}\widehat{\pr}\{\psi_{\tau,\acw}(V;\widehat{\mu}_{1},\widehat{\mu}_{0},\widehat{q},\widehat{r})\}+o_{p}(N^{-1/2})\\
 & =\widehat{\pr}\{\psi_{\tau,\acw}(V;\widehat{\mu}_{1},\widehat{\mu}_{0},\widehat{q},\widehat{r})\}+o_{p}(N^{-1/2}),
\end{align*}
where $\E\left\{ \partial\psi_{\tau,\acw}(V;\mu_{1},\mu_{0},q,r)/\partial\tau\right\} =-1$.
Moreover, we can show 
\begin{align}
 & \widehat{\pr}\{\psi_{\tau,\acw}(V;\widehat{\mu}_{1},\widehat{\mu}_{0},\widehat{q},\widehat{r})\}\nonumber\\
 & =(\widehat{\pr}-\pr)\{\psi_{\tau,\acw}(V;\widehat{\mu}_{1},\widehat{\mu}_{0},\widehat{q},\widehat{r})\}+\pr\{\psi_{\tau,\acw}(V;\widehat{\mu}_{1},\widehat{\mu}_{0},\widehat{q},\widehat{r})\}\nonumber\\
 & =\widehat{\pr}\{\psi_{\tau,\acw}(V;\mu_{1},\mu_{0},q,r^{*})\}\nonumber\\
 & +\pr\{\psi_{\tau,\acw}(V;\widehat{\mu}_{1},\widehat{\mu}_{0},\widehat{q},\widehat{r})-\psi_{\tau,\acw}(V;\mu_{1},\mu_{0},q,r^{*})\}\label{eq:second_term}\\
 & + (\widehat{\pr}-\pr)\{\psi_{\tau,\acw}(V;\widehat{\mu}_{1},\widehat{\mu}_{0},\widehat{q},\widehat{r})-\psi_{\tau,\acw}(V;\mu_{1},\mu_{0},q,r^{*})\}\label{eq:third_term},
\end{align}
where the third term (\ref{eq:third_term}) is $o_{p}(N^{-1/2})$ under Assumption \ref{assum:rate-conditions}
and \ref{assum:regularity-eff} a). Even if the Donsker condition in Assumption \ref{assum:regularity-eff} a) is not met, the cross-fitting procedure in \cite{chernozhukov2018double} can be used to assure that (\ref{eq:third_term}) is negligible. We now show that the second term (\ref{eq:second_term}) is a small order term:
\begin{align*}
 & P(R=1)\pr\{\psi_{\tau,\acw}(V;\widehat{\mu}_{1},\widehat{\mu}_{0},\widehat{q},\widehat{r})-\psi_{\tau,\acw}(V;\mu_{1},\mu_{0},q,r^{*})\}\\
 & =-\pr[R\{\widehat{\mu}_{0}(X)-\mu_{0}(X)\}]\\
 & -\pr\left[\frac{\widehat{q}(X)\widehat{r}(X)P(R=0\mid X)+\widehat{q}(X)P(R=1\mid X)\{1-\pi_{A}(X)\}}{\widehat{r}(X)+\{1-\pi_{A}(X)\}\widehat{q}(X)}\left\{ \mu_{0}(X)-\widehat{\mu}_{0}(X)\right\} \right]\\
 & =-\pr\left[P(R=0\mid X)q(X)\{\widehat{\mu}_{0}(X)-\mu_{0}(X)\}\right]\\
 & +\pr\left[P(R=0\mid X)\widehat{q}(X)\frac{\widehat{r}(X)+q(X)\{1-\pi_{A}(X)\}}{\widehat{r}(X)+\{1-\pi_{A}(X)\}\widehat{q}(X)}\left\{ \widehat{\mu}_{0}(X)-\mu_{0}(X)\right\} \right]\\
 & =\pr[P(R=0\mid X)\{\widehat{q}(X)-q(X)\}\{\widehat{\mu}_{0}(X)-\mu_{0}(X)\}]\\
 & +\pr\left[P(R=0\mid X)\widehat{q}(X)\left\{ \frac{\widehat{r}(X)+q(X)\{1-\pi_{A}(X)\}}{\widehat{r}(X)+\{1-\pi_{A}(X)\}\widehat{q}(X)}-1\right\} \left\{ \widehat{\mu}_{0}(X)-\mu_{0}(X)\right\} \right]\\
 & =\pr[P(R=0\mid X)\{\widehat{q}(X)-q(X)\}\{\widehat{\mu}_{0}(X)-\mu_{0}(X)\}]\\
 & +\pr\left[P(R=0\mid X)\frac{\widehat{q}(X)\{1-\pi_{A}(X)\}}{\widehat{r}(X)+\{1-\pi_{A}(X)\}\widehat{q}(X)}\{q(X)-\widehat{q}(X)\}\left\{ \widehat{\mu}_{0}(X)-\mu_{0}(X)\right\} \right].
\end{align*}
Under Assumption \ref{assum:exchange_delta}, $P(R=0\mid X)>0$ and
$\widehat{q}(X)\{1-\pi_{A}(X)\}/\{\widehat{r}(X)+(1-\pi_{A})\widehat{q}(X)\}$
is bounded by 1. Therefore, by Cauchy-Schwarz inequality, $\widehat{\tau}_{\acw}-\tau-\widehat{\pr}\{\psi_{\tau,\acw}(V;\mu_{1},\mu_{0},q,r^{*})\}$
is bounded by $C^{2}\|\widehat{q}(X)-q(X)\|\|\widehat{\mu}_{0}(X)-\mu_{0}(X)\|$, which is $o_{p}(N^{-1/2})$ under Assumption \ref{assum:regularity-seives}. Therefore, the bias is asymptotically negligible, and the influence
function $\psi_{\tau,\acw}(V;\widehat{\mu}_{1},\widehat{\mu}_{0},\widehat{q},\widehat{r})$
can be used for obtaining a robust variance estimation, thus $\widehat{\bV}_{\tau}=N^{-1}\sum_{i\in\mathcal{R}\cup\mathcal{E}}\psi_{\tau,\acw}^{2}(V;\widehat{\mu}_{1},\widehat{\mu}_{0},\widehat{q},\widehat{r})$,
which is consistent for $\E\{\psi_{\tau,\acw}^{2}(V;\mu_{1},\mu_{0},q,r^{*})\}$, and $\widehat{\tau}_{\acw}$ attains
the semiparametric efficiency when $r^*(X)=r(X)$.  When the propensity score $\pi_A(X)$ is unknown and is estimated by $\widehat{\pi}_A(X)$, the second term (\ref{eq:second_term}) becomes
\begin{align*}
 & P(R=1)\pr\{\psi_{\tau,\acw}(V;\widehat{\pi}_A,\widehat{\mu}_{1},\widehat{\mu}_{0},\widehat{q},\widehat{r})-\psi_{\tau,\acw}(V;{\pi}_A,\mu_{1},\mu_{0},q,r^{*})\}\\
 & = \pr\left[
    \frac{\{\widehat{\pi}_A(X)-\pi_A(X)\}\{\widehat{\mu}_1(X)-\mu_1(X)\}}{\widehat{\pi}_A(X)}
    \right]\\
 & + \pr[P(R=0\mid X)\{\widehat{q}(X)-q(X)\}\{\widehat{\mu}_{0}(X)-\mu_{0}(X)\}]\\
 & +\pr\left[P(R=0\mid X)\frac{\widehat{q}(X)\{1-\pi_{A}(X)\}}{\widehat{r}(X)+\{1-\pi_{A}(X)\}\widehat{q}(X)}\{q(X)-\widehat{q}(X)\}\left\{ \widehat{\mu}_{0}(X)-\mu_{0}(X)\right\} \right],
\end{align*}
where the first term is controlled by the estimation error of $(\widehat{\pi}_A(X),\widehat{\mu}_1(X))$, and the last two terms are controlled by the estimation error of $(\widehat{q}(X),\widehat{\mu}_0(X))$. Thus, the estimator $\widehat{\tau}_{\acw}$ is consistent if $\|\widehat{\pi}_A(X)-\pi_A(X)\|\|\widehat{\mu}_{1}(X)-\mu_{1}(X)\|$ and $\|\widehat{q}(X)-q(X)\|\|\widehat{\mu}_{0}(X)-\mu_{0}(X)\|$ are $o_p(N^{-1/2})$.

\subsection{Proof of Lemma \ref{lem:selection-consistency}}\label{subsec:proof_lemma}
For $i\in\mathcal{E}$, denote the true subject-level bias parameter by $b_{i,0}=b_0(X_i)=\E(Y_{i}\mid X_{i},R_i=0)-\E(Y_{i}\mid X_{i},R_{i}=1, A_i = 0)$. By the iterated expectation, we have
\begin{align*}
b_{i,0} &=\E(Y_{i}\mid X_{i},R_i=0)-\E(Y_{i}\mid X_{i},R_{i}=1, A_i = 0)\\
&=\mu_{0,\mathcal{E}}(X_i) - \mu_{0}(X_i).
\end{align*}
Therefore, for $i\in\mathcal{E}$, $b_{i,0}$ can be characterized
by $\widehat{b}_i=\widehat{\mu}_{0,\mathcal{E}}(X_i) - \widehat{\mu}_{0}(X_i)$. Let the penalized estimators $\tilde{b}$ be the solution to
\begin{align*}
\tilde{b}&=\arg\min_{b}\left\{ (\widehat{b}-b)^{\T}\widehat{\Sigma}_{b}^{-1}(\widehat{b}-b)+\lambda_{N}\sum_{i\in\rwd}p(|b_{i}|)\right\} \\
   &=\arg\min_{b}\left\{ (\widehat{\Gamma}^\T\widehat{b}-\widehat{\Gamma}^\T b)^{\T}(\widehat{\Gamma}^\T\widehat{b}-\widehat{\Gamma}^\T b)+\lambda_{N}\sum_{i\in\rwd}w_i |b_i|\right\}\\
   &=\arg\min_{b}\left\{ (\widehat{b}^*-\widehat{\Gamma}^\T b)^{\T}(\widehat{b}^*-\widehat{\Gamma}^\T b)+\lambda_{N}\sum_{i\in\rwd}
   \frac{|b_i|}{|\widehat{b}_i|^\nu}
   \right\},
\end{align*}
where $\widehat{\Sigma}_b^{-1}$ is the variance of $\widehat{b}$, $\widehat{\Sigma}_b^{-1} = \widehat{\Gamma} \widehat{\Gamma}^\T$, and $\widehat{b}^* = \widehat{\Gamma}^\T\widehat{b}$. Without loss of generality, we assume the first $|\mathcal{A}^c|$ of $b_0$ are non-zero and the rest of the entries are zero, i.e., $b_0 = (b_{10}, 0)$. Then $\widehat{\Sigma}_b^{-1}= \widehat{\Gamma} \widehat{\Gamma}^\T$ can be expressed in a block-wise form as follows:
$$
\widehat{\Sigma}_b^{-1}=\widehat{\Gamma} \widehat{\Gamma}^\T = 
\begin{pmatrix}
    \widehat{\Gamma}_1\\
    \widehat{\Gamma}_2
\end{pmatrix}
(\widehat{\Gamma}_1^\T, \widehat{\Gamma}_2^\T) = 
\begin{pmatrix}
  \widehat{\Gamma}_{1}\widehat{\Gamma}_{1}^\T & \widehat{\Gamma}_{1}\widehat{\Gamma}_{2}^\T\\
\widehat{\Gamma}_{2}\widehat{\Gamma}_{1}^\T & \widehat{\Gamma}_{2}\widehat{\Gamma}_{2}^\T
\end{pmatrix}.
$$
It follows from the Karush–Kuhn–Tucker conditions that $\tilde{b}$ is the unique solution if
$$
\begin{cases}
    \widehat{\gamma}_i (\widehat{b}^*-\widehat{\Gamma}^\T \tilde{b})= \lambda_N w_i \text{sgn}(\tilde{b}_{i,0}), &  i \notin \mathcal{A},\\
    |\widehat{\gamma}_i (\widehat{b}^*-\widehat{\Gamma}^\T \tilde{b})| < \lambda_N w_i, & i \in \mathcal{A},
\end{cases}
$$
where $\widehat{\gamma}_i$ is the $i$-th row of $\widehat{\Gamma}$. Let $w_{N1}=(w_i\text{sgn}(b_{i,0}), i\notin\mathcal{A})^\T$ and
\begin{align*}
   \tilde{b}_1 & = (\widehat{\Gamma}_1 \widehat{\Gamma}_1^\T)^{-1} (\widehat{\Gamma}_1 \widehat{b}^* - \lambda_N w_{N1}) \\
   & = 
(\widehat{\Gamma}_1 \widehat{\Gamma}_1^\T)^{-1}\widehat{\Gamma}_1 \widehat{b}^*
- (\widehat{\Gamma}_1 \widehat{\Gamma}_1^\T)^{-1} \lambda_N w_{N1}\\ 
& = (\widehat{\Gamma}_1 \widehat{\Gamma}_1^\T)^{-1}\widehat{\Gamma}_1
\widehat{\Gamma}^\T \{\widehat{\mu}_{0,\mathcal{E}}(X) - \widehat{\mu}_0(X)\} - (\widehat{\Gamma}_1 \widehat{\Gamma}_1^\T)^{-1} \lambda_N w_{N1} \\
& = b_{10} + 
(\widehat{\Gamma}_1 \widehat{\Gamma}_1^\T)^{-1}\widehat{\Gamma}_1
\widehat{\Gamma}^\T \{\widehat{\mu}_{0,\mathcal{E}}(X) - \widehat{\mu}_0(X) - b_0\} - (\widehat{\Gamma}_1 \widehat{\Gamma}_1^\T)^{-1}\lambda_N w_{N1}.
\end{align*}
By Thereom 1 of \cite{huang2008adaptive} and Theorem 2.1 of \cite{lin2009adaptive}, the Karush–Kuhn–Tucker condition holds for $\tilde{b} = (\tilde{b}_1, 0)$ if $\text{sgn}(\tilde{b}_1) = \text{sgn}(b_{10})$. Thus, since $\widehat{\Gamma}^\T \tilde{b} = \widehat{\Gamma}_1^\T \tilde{b}_1$ for $\tilde{b}=(\tilde{b}_1,0)$, we have $\pr(\mathcal{A}=\tilde{\mathcal{A}})$ if
$$
\begin{cases}
 \text{sgn}(\tilde{b}_1) = \text{sgn}(b_{10}),\\
|\widehat{\gamma}_i(\widehat{b}^* -\widehat{\Gamma}_1^\T\tilde{b}_1)| < 
\lambda_N w_i, & i \in \mathcal{A}.
\end{cases}
$$
Since $\{|b_{10} - \tilde{b}_1|<|b_{10}|\}\subset \{\text{sgn}(\tilde{b}_1)=\text{sgn}(b_{10})\}$, we aim to prove that
\begin{align*}
\pr(\mathcal{A}\neq \tilde{\mathcal{A}}) &\leq 
\pr(|b_{i, 10} - \tilde{b}_{i,1}| \geq |b_{i, 10}|, \exists i \notin \mathcal{A}) + 
\pr(|\widehat{\gamma}_i(\widehat{b}^* -\Gamma_1^\T\tilde{b}_1)| \geq 
\lambda_N w_i, \exists i \in \mathcal{A})\\
    &=
    \pr(E_1) + \pr(E_2) \rightarrow 0.
\end{align*}
First, we have $\pr(E_1) = \pr(\cup_{i\notin \mathcal{A}} E_{1i})$, where $E_{1i}=\{|b_{i,10} - \tilde{b}_{i,1}|\geq |b_{i,10}|\}$ for $i\notin\mathcal{A}$ and
\begin{align*}
    \pr(E_{1i}) & = \pr(|b_{i,10} - \tilde{b}_{i,1}|\geq |b_{i,10}|) \\
    &\leq 
    \pr(|e_i^\T(\widehat{\Gamma}_1 \widehat{\Gamma}_1^\T)^{-1}\widehat{\Gamma}_1
\widehat{\Gamma}^\T \{\widehat{\mu}_{0,\mathcal{E}}(X) - \widehat{\mu}_0(X) - b_0\}|\geq |b_{i,0}|/2) \\
& + \pr(|e_i^\T(\widehat{\Gamma}_1 \widehat{\Gamma}_1^\T)^{-1} \lambda_N w_{N1}|\geq |b_{i,0}|/2),
\end{align*}
where $e_i$ is a zero-valued vector except its $i$-th entry. Putting these terms together, we have
\begin{align}
    \pr(E_1)& =  \pr(\cup_{i\notin \mathcal{A}} E_{1i}) \nonumber \\
    & \leq 
    \pr(|e_i^\T(\widehat{\Gamma}_1 \widehat{\Gamma}_1^\T)^{-1}\widehat{\Gamma}_1
\widehat{\Gamma}^\T \{\widehat{\mu}_{0,\mathcal{E}}(X) - \widehat{\mu}_0(X) - b_0\}|\geq |b_{i,0}|/2, \exists i \notin \mathcal{A}) \nonumber \\
    & + \pr(|e_i^\T(\widehat{\Gamma}_1 \widehat{\Gamma}_1^\T)^{-1} \lambda_N w_{N1}|\geq |b_{i,0}|/2, \exists i \notin \mathcal{A})\nonumber \\
    & \leq 
    \pr(a_N \tau_2\tau_1^{-1} \max_{i\not\in \mathcal{A}}|\widehat{\mu}_{0,\mathcal{E}}(X_i) - \widehat{\mu}_0(X_i) - b_0 |\geq |a_N b_{\min}|/2) \label{E1:term1}\\
    & + P\left(\left|\frac{\tau_2\lambda_N }{|\widehat{\mu}_{0,\mathcal{E}}(X_i)-\widehat{\mu}_{0}(X_i)|^\nu}\right|\geq |b_{\min}|/2,  \exists i \notin \mathcal{A}
    \right),\label{E1:term3}
\end{align}
where the second inequality holds under Assumption \ref{assum:selection-regularity}(ii).

For term (\ref{E1:term1}), we have $a_N\max_i\{\widehat{\mu}_{0}(X_i) -\mu_0(X_i)\}=O_p(1)$ and $a_N\max_i\{\widehat{\mu}_{0,\mathcal{E}}(X_i) -\mu_{0,\mathcal{E}}(X_i)\}=O_p(1)$ by Assumption \ref{assum:selection-regularity}(i) and $b_{i,0} = \mu_{0,\mathcal{E}}(X_i) - \mu_{0}(X_i)$ by definition. Therefore, $\pr(a_N \tau_2\tau_1^{-1} \max_{i\in \mathcal{E}}|\widehat{\mu}_{0,\mathcal{E}}(X_i) - \widehat{\mu}_0(X_i) - b_0 |\geq |a_N b_{\min}|/3)\rightarrow 0$ for any $i\notin \mathcal{A}$ as $a_N b_{\min}\rightarrow \infty$ under Assumption \ref{assum:selection-regularity}(iii).

For term (\ref{E1:term3}), we have $|\widehat{\mu}_{0,\mathcal{E}}(X_i)-\widehat{\mu}_{0}(X_i)|\overset{p}{\rightarrow} b_{i,0}\leq b_{\min}$ for $i\notin\mathcal{A}$ by Assumption \ref{assum:selection-regularity}(i), and therefore
$$
P\left(\left|\frac{\lambda_N \tau_2}{|\widehat{\mu}_{0,\mathcal{E}}(X_i)-\widehat{\mu}_{0}(X_i)|^\nu}\right|\geq |b_{\min}|/3
    \right)
    \leq 
P\left(\left|\frac{\lambda_N\tau_2}{|b_{\min}|^\nu}\right|\geq |b_{\min}|/3
\right) \rightarrow 0,
$$
which holds as $\lambda_N/b_{\min}^{\nu+1}\rightarrow 0$  under Assumption \ref{assum:selection-regularity}(iv). Thus, we prove that $\pr(E_1)\rightarrow 0$. 

Next, we have that $\pr(E_{2})= \pr(\cup_{i\in \mathcal{A}} E_{2i})$, where $E_{2i}=\{|\widehat{\gamma}_i(\widehat{b}^* -\widehat{\Gamma}_1^\T\tilde{b}_1)| \geq \lambda_N w_i\}$ for $i\in\mathcal{A}$ and 
\begin{align*}
    \widehat{b}^* -\widehat{\Gamma}_1^\T\tilde{b}_1 & = 
\widehat{b}^* -\widehat{\Gamma}_1^\T(\widehat{\Gamma}_1 \widehat{\Gamma}_1^\T)^{-1} (\widehat{\Gamma}_1 \widehat{b}^* - \lambda_N w_{N1}) \\
& = 
(I_{N} - H)\widehat{b}^* + \widehat{\Gamma}_1^\T(\widehat{\Gamma}_1 \widehat{\Gamma}_1^\T)^{-1}\lambda_N w_{N1} \\
& = (I_{N} - H)\widehat{\Gamma}^\T \{\widehat{\mu}_{0,\mathcal{E}}(X) - \widehat{\mu}_0(X)\} + \widehat{\Gamma}_1^\T(\widehat{\Gamma}_1 \widehat{\Gamma}_1^\T)^{-1}\lambda_N w_{N1},
\end{align*}
where $H = \Gamma_1^\T(\widehat{\Gamma}_1 \widehat{\Gamma}_1^\T)^{-1} \widehat{\Gamma}_1$. This is a variant of Theorem 1 of \cite{huang2008adaptive}. Thus, we can show
\begin{align*}
    \pr(E_{2i})&  = 
    P\{|\widehat{\gamma}_i(\widehat{b}^* -\widehat{\Gamma}_1^\T\tilde{b}_1)| \geq \lambda_N w_i\}\\
    &\leq \pr(|\widehat{\gamma}_i(I_{N} - H)\widehat{\Gamma}^\T \{\widehat{\mu}_{0,\mathcal{E}}(X_i) - \widehat{\mu}_0(X_i)\}| \geq \lambda_N w_i/2) \\
    & + \pr(|\widehat{\gamma}_i\widehat{\Gamma}_1^\T(\widehat{\Gamma}_1 \widehat{\Gamma}_1^\T)^{-1}\lambda_N w_{N1}|\geq \lambda_N w_i/2).
\end{align*}
Putting these terms together, we have
\begin{align}
    \pr(E_2)& = \pr(\cup_{i\in \mathcal{A}} E_{2i})\nonumber\\
    & \leq \pr(|\widehat{\gamma}_i(I_{N} - H)\widehat{\Gamma}^\T \{\widehat{\mu}_{0,\mathcal{E}}(X) - \widehat{\mu}_0(X)\}| \geq \lambda_N w_i/2, \exists i \in \mathcal{A})\nonumber \\
    & + \pr(|\widehat{\gamma}_i\widehat{\Gamma}_1^\T(\widehat{\Gamma}_1 \widehat{\Gamma}_1^\T)^{-1}\lambda_N w_{N1}|\geq \lambda_N w_i/2, \exists i \in \mathcal{A})\nonumber \\
    & \leq \pr(\max_{i\in\mathcal{A}}|\widehat{\mu}_{0,\mathcal{E}}(X_i) - \widehat{\mu}_0(X_i)|/\tau_1 \geq \lambda_N w_i/2)\label{E2:term1} \\
    & + \pr(\tau_2\tau_1^{-1}\lambda_N \|w_{N1}\|\geq \lambda_N w_i/2, \exists i \in \mathcal{A}),\label{E2:term3}
\end{align}
where the second inequality holds under Assumption \ref{assum:selection-regularity}(ii).

For term (\ref{E2:term1}), we have $a_N\max_{i\in\mathcal{A}}\{\widehat{\mu}_{0,\mathcal{E}}(X_i)-\widehat{\mu}_{0}(X_i)\}=O_p(1)$ as $b_{i,0}=0$, and therefore $\pr(|a_N\{\mu_{0,\mathcal{E}}(X_i) - \widehat{\mu}_0(X_i)\}|/\tau_1 \geq a_N\lambda_N w_i/3) \rightarrow \infty$ for any $i\in \mathcal{A}$ as $\lambda_N a_N^{\nu}\rightarrow\infty$ under Assumption \ref{assum:selection-regularity}(iv).


For term (\ref{E2:term3}), we have
\begin{align*}
    &\pr(\tau_2\tau_1^{-1}\lambda_N \|w_{N1}\|\geq \lambda_N w_i/3, \exists i \in \mathcal{A}) 
    \\
    & \leq \pr\left(
    \frac{\tau_2\tau_1^{-1}\lambda_N}{b_{\min}^\nu}
    \geq \frac{\lambda_N a_N^\nu}{3|a_N\{\widehat{\mu}_{0,\mathcal{E}}(X_i)-\widehat{\mu}_{0}(X_i)\}|^\nu}\right) \rightarrow 0,
\end{align*}
which holds as $a_N\max_{i\in\mathcal{A}}\{\widehat{\mu}_{0,\mathcal{E}}(X_i)-\widehat{\mu}_{0}(X_i)\}=O_p(1)$ and $a_N b_{\min} \rightarrow \infty$ under Assumptions \ref{assum:selection-regularity}(i) and (iii). Therefore, we prove that $\pr(E_2)\rightarrow 0$. 

To sum up, we can show the selection consistency, that is $\pr(\tilde{\mathcal{A}}=\mathcal{A}) = 1 - \pr(\tilde{\mathcal{A}}\neq \mathcal{A}) \rightarrow 1$, which completes our proof.

\subsection{Proof of Theorem \ref{thm:acw-final-point}}
\textit{Consistency}: Denote the influence function for the data-adaptive
integrative estimator by
\begin{align*}
 & \psi_{\tau,\acw}^{\lasso}(V;\mu_{1},\mu_{0},q,r_{b})=\frac{R}{P(R=1)}\left[\{\mu_{1}(X)-\mu_{0}(X)-\tau\}+\frac{A\{Y-\mu_{1}(X)\}}{\pi_{A}(X)}\right]\\
 & -\frac{(1-R)\mathds{1}(b_{0}=0)}{P(R=1)}\frac{q(X)r_b(X)\{Y-\mu_{0}(X)\}}{r_b(X)P(b_{0}=0\mid X, R=0)+\{1-\pi_{A}(X)\}q(X)}\\
 & -\frac{R(1-A)}{P(R=1)}\frac{q(X)\left\{ Y-\mu_{0}(X)\right\}}{r_b(X)P(b_{0}=0\mid X, R=0)+\{1-\pi_{A}(X)\}q(X)},
\end{align*}
where $r_b(X)=\text{var}(Y\mid X,R=1, A=0)/\text{var}(Y\mid X,R=0,b_{0}=0\}$. Under the conditions in Theorem \ref{thm:acw-final-point}, $\E\{\psi_{\tau,\acw}^{\lasso}(V;\mu_{1},\mu_{0},q,r_{b})\}=0$.
Next, the consistency of $\widehat{\tau}_{\acw}^{\lasso}$ can be
justified similar to Section \ref{subsec:Proof-of-Theorem-rate_robust}
by showing
\begin{align*}
 & \pr\left\{ \psi_{\tau,\acw}^{\lasso}(V;\widehat{\mu}_{1},\widehat{\mu}_{0},\widehat{q},\widehat{r}_b)-\psi_{\tau,\acw}^{\lasso}(V;\mu_{1},\mu_{0},q,r^{*}_b)\right\} \\
= & \pr[P(R=0\mid X)\{\widehat{q}(X)-q(X)\}\{\widehat{\mu}_{0}(X)-\mu_{0}(X)\}]\\
 & +\pr\left[P(R=0\mid X)\frac{\widehat{q}(X)\{1-\pi_{A}(X)\}}{\widehat{r}_b(X)\widehat{P}(\tilde{b}=0\mid X, R=0)+\{1-\pi_{A}(X)\}\widehat{q}(X)}\{q(X)-\widehat{q}(X)\}\left\{ \widehat{\mu}_{0}(X)-\mu_{0}(X)\right\} \right],
\end{align*}
which is bounded by $C^{2}\|\widehat{q}(X)-q(X)\|\|\widehat{\mu}_{0}(X)-\mu_{0}(X)\|=o_{p}(N^{-1/2})$ under Assumption \ref{assum:regularity-seives}. 

\textit{Efficiency}: From \citet{cao2009improving}, we know that
the trial-only efficient influence function is $\psi_{\tau,\aipw}(V;\mu_{1},\mu_{0})$,
and the asymptotic variance for $\widehat{\tau}_{\aipw}$ is $V_{\tau,\aipw}=\E\{\psi_{\tau,\aipw}^{2}(V;\mu_{1},\mu_{0})\}$,
where
\begin{align*}
\psi_{\tau,\aipw}(V;\mu_{1},\mu_{0}) & =\frac{R}{P(R=1)}\{\mu_{1}(X)-\mu_{0}(X)-\tau\}\\
 & +\frac{R}{P(R=1)}\left[\frac{A\{Y-\mu_{1}(X)\}}{\pi_{A}(X)}-\frac{(1-A)\{Y-\mu_{0}(X)\}}{1-\pi_{A}(X)}\right].
\end{align*}
Compared the asymptotic variance $\bV_{\tau,\aipw}$ to $\bV_{\tau,\acw}^{\lasso}$,
we have 
\begin{align}
V_{\tau,\aipw}-V_{\tau,\acw}^{\lasso} & =\E\{\psi_{\tau,\aipw}^{2}(V;\mu_{1},\mu_{0})\}-\E\{\psi_{\tau,\acw}^{\lasso}(V;\mu_{1},\mu_{0},q,r^{*})\}^{2}\nonumber \\
 & =\E\{\psi_{\tau,\acw}^{\lasso}(V;\mu_{1},\mu_{0},q,r^{*})+\psi_{\tau,\aipw}(V;\mu_{1},\mu_{0})-\psi_{\tau,\acw}^{\lasso}(V;\mu_{1},\mu_{0},q,r^{*})\}^{2}\nonumber \\
 & -\E\{\psi_{\tau,\acw}^{\lasso}(V;\mu_{1},\mu_{0},q,r^{*})\}^{2}\nonumber \\
 & =\E\{\psi_{\tau,\aipw}(V;\mu_{1},\mu_{0})-\psi_{\tau,\acw}^{\lasso}(V;\mu_{1},\mu_{0},q,r^{*})\}^{2}\label{eq:V_aipw-V_acw}\\
 & +2\E\left[\psi_{\tau,\acw}^{\lasso}(V;\mu_{1},\mu_{0},q,r^{*})\{\psi_{\tau,\aipw}(V;\mu_{1},\mu_{0})-\psi_{\tau,\acw}^{\lasso}(V;\mu_{1},\mu_{0},q,r^{*})\}\right].\nonumber 
\end{align}
By pathwise derivative, we can show that $\{ \psi_{\tau,\aipw}(V;\mu_{1},\mu_{0})-\psi_{\tau,\acw}^{\lasso}(V;\mu_{1},\mu_{0},q,r^{*})\}$ is orthogonal to $\mathcal{T}$. Also, we can verify
that $\psi_{\tau,\acw}^{\lasso}(V;\mu_{1},\mu_{0},q,r^{*})$
belongs to the updated tangent space $\mathcal{T} = \mathcal{T}_1 + \mathcal{T}_2 + \mathcal{T}_3 + \mathcal{T}_4^*$ where
\begin{align*}
\mathcal{T}_4^* & = \mathcal{T}_{41}\cap \mathcal{T}_{42}^*=\left\{ \Gamma(Y, X, R, A):\int\Gamma(y, X, R, A)f(y\mid X, R, A)dy=0\right\} \\
&\cap \left\{ \Gamma(Y, X, R, A):
\E\left[\left\{
\frac{(1-R)\mathds{1}(b_0=0)Y}{P(R=0\mid X)}- 
\frac{R(1-A)Y}{P(R=1, A=0\mid X)}\right\}\Gamma(Y, X, R, A)\mid X\right]=0\right\},
\end{align*}
since its components satisfy
\begin{align*}
 & \{R-P(R=1\mid X)\}\{\mu_{1}(X)-\mu_{0}(X)-\tau\}\in\mathcal{T}_{2},\\
 & P(R=1\mid X)\{\mu_{1}(X)-\mu_{0}(X)-\tau\}\in\mathcal{T}_{1},\quad \frac{RA\{Y-\mu_{1}(X)\}}{\pi_{A}(X)}\in\mathcal{T}_{4}^*,\\
 & \frac{R(1-A)q(X)\left\{ Y-\mu_{0}(X)\right\}}{r_b^{*}(X)P(b_{0}=0\mid X, R=0)+\{1-\pi_{A}(X)\}q(X)}  \\
 &+ \frac{(1-R)\mathds{1}(b_{0}=0)q(X)r_b^{*}(X)\{Y-\mu_{0}(X)\}}{r_b^{*}(X)P(b_{0}=0\mid X, R=0)+\{1-\pi_{A}(X)\}q(X)}\in\mathcal{T}_{4}^*.
\end{align*}
Thus, $\psi_{\tau,\acw}^{\lasso}(V;\mu_{1},\mu_{0},q,r^{*})\in \mathcal{T}$, and the second term $$\E\left[\psi_{\tau,\acw}^{\lasso}(V;\mu_{1},\mu_{0},q,r^{*})\{\psi_{\tau,\aipw}(V;\mu_{1},\mu_{0})-\psi_{\tau,\acw}^{\lasso}(V;\mu_{1},\mu_{0},q,r^{*})\}\right]=0.$$
In addition, for the first term (\ref{eq:V_aipw-V_acw}), we have
\begin{align*}
 & P^{2}(R=1)\E\{\psi_{\tau,\aipw}(V;\mu_{1},\mu_{0})-\psi_{\tau,\acw}^{\lasso}(V;\mu_{1},\mu_{0},q,r_b^{*})\}^{2}\\
= & \E\left[R(1-A)\left\{ \frac{q(X)}{G^*(X)}-\frac{1}{1-\pi_{A}(X)}\right\} ^{2}\{Y-\mu_{0}(X)\}^{2}\right]\\
 & +\E\left[(1-R)\mathds{1}(b_{0}=0)\left\{ \frac{r_b^*(X)q(X)}{G^*(X)}\right\} ^{2}\{Y-\mu_{0}(X)\}^{2}\right]\\
 = & \E\left[P(R=1\mid X)\{1-\pi_{A}(X)\}\left\{ \frac{q(X)}{G^*(X)}-\frac{1}{1-\pi_{A}(X)}\right\} ^{2}\text{var}(Y\mid X,R=1, A=0)\right]\\
 & +\E\left[P(R=0\mid X) \mathds{1}(b_{0}=0)  \left\{\frac{r_b^*(X)q(X)}{G^*(X)}\right\}^{2}\text{var}(Y\mid X,R=0,b_0=0)\right]\\
= & \E\left[P(R=1\mid X)\{1-\pi_{A}(X)\}\left\{ \frac{q(X)}{G^*(X)}-\frac{1}{1-\pi_{A}(X)}\right\} ^{2}\text{var}(Y\mid X,R=1, A=0)\right]\\
 & +\E\left[P(R=1\mid X) \mathds{1}(b_{0}=0)  \frac{r_b^*(X)^2q(X)}{r_b(X) G^*(X)^{2}}\text{var}(Y\mid X,R=1, A=0)\right]\\
= & \E\left[\frac{\pr(R=1\mid X)r^{*}_b(X)\mathds{1}(b_{0}=0)}{G^*(X)\{1-\pi_A(X)\}}
\frac{r_b^*(X)}{r_b(X)}
\frac{G(X)}{G^*(X)}
\text{var}(Y\mid X,R=1, A=0)\right]\geq0,
\end{align*}
where $G(X)=r_b(X)P(b_{0}=0\mid X, R=0)+\{1-\pi_{A}(X)\}q(X)$ and $G(X)$ with $r_b(X)$ being replaced by $r_b^{*}(X)$ for notation simplicity. Hence, it completes our proof if $r^*(X)=r(X)$. 

\subsection{Proof of Theorem \ref{thm:acw-final-CI}}\label{subsec:proofCI}
The proof of Theorem \ref{thm:acw-final-CI} is presented in three
folds. (i) First, we show
that the estimator $\widehat{\tau}_{\acw}^{\lasso}(\mathcal{A})$
with the subset $\mathcal{A}$ known apriori is consistent for $\tau$.
(ii) Next, the selection consistency established in Lemma \ref{lem:selection-consistency}
ensures that the penalized estimator $\widehat{\tau}_{\acw}^{\lasso}(\tilde{\mathcal{A}})$
is asymptotically equivalent to the estimator $\widehat{\tau}_{\acw}^{\lasso}(\mathcal{A})$. Therefore, the effect of the estimation uncertainty of $\tilde{\mathcal{A}}$
on the penalized estimator $\widehat{\tau}_{\acw}^{\lasso}$ can be
neglected. (iii) Finally, the consistency and asymptotic normality of
$\widehat{\tau}_{\acw}^{\lasso}$ is established under Assumption
\ref{assum:rate-conditions}, and the confidence interval defined in (\ref{eq:CI_acw_lasso})
is valid over any chosen subset $\tilde{\mathcal{A}}$, which completes
the proof of Theorem \ref{thm:acw-final-CI}. We now provide the details of the proof.

In Lemma \ref{lem:acw_lass_fixed}, we first establish the
asymptotic distribution for $\widehat{\tau}_{\acw}^{\lasso}(\mathcal{A})$
with fixed $\mathcal{A}$.

\begin{lemma}\label{lem:acw_lass_fixed}

Let $\mathcal{A}=\{i:b_{i,0}=0\}$ and the augmented inverse probability
weighting estimator with the oracle selection subset $\mathcal{A}$
be
\begin{align*}
\widehat{\tau}_{\acw}^{\lasso}(\mathcal{A}) & =N_{\mathcal{R}}^{-1}\sum_{i\in\mathcal{R}\cup\mathcal{E}}R_{i}\left[\frac{A_{i}\{Y_{i}-\widehat{\mu}_{1}(X_{i})\}}{\pi_{A}(X_{i})}+\widehat{\mu}_{1}(X_{i})-\widehat{\mu}_{0}(X_{i})\right]\\
 & -N_{\mathcal{R}}^{-1}\sum_{i\in\mathcal{R}\cup\mathcal{E}}\frac{R_{i}(1-A_{i})\widehat{q}(X_{i})}{\widehat{r}_b(X_i)P(b_{i,0}=0\mid X, R=0)+\{1-\pi_{A}(X_{i})\}\widehat{q}(X_{i})}\left\{ Y_{i}-\widehat{\mu}_{0}(X_{i})\right\} \\
 & -N_{\mathcal{R}}^{-1}\sum_{i\in\mathcal{R}\cup\mathcal{E}}\frac{(1-R_{i})\mathds{1}({b}_{i,0}=0)\widehat{q}(X_{i})\widehat{r}_b(X_i)}{\widehat{r}_b(X_i)P(b_{i,0}=0\mid X, R=0)+\{1-\pi_{A}(X_{i})\}\widehat{q}(X_{i})}\left\{ Y_{i}-\widehat{\mu}_{0}(X_{i})\right\} .
\end{align*}
Under Assumptions \ref{a:ign} and \ref{assum:rate-conditions}, we
have $N^{1/2}\{\widehat{\tau}_{\acw}^{\lasso}(\mathcal{A})-\tau\}\rightarrow N\{0,\bV_{\tau}^{\lasso}(\mathcal{A})\}$
when $N\rightarrow\infty$. 

\end{lemma}
Next, we show that the uncertainty originated from the selected subset $\tilde{\mathcal{A}}$ is negligible in Lemma \ref{lem:acw_lass_selection}.

\begin{lemma}\label{lem:acw_lass_selection}

Under Assumptions \ref{a:ign} and \ref{assum:rate-conditions}, we
have $N^{1/2}\{\widehat{\tau}_{\acw}^{\lasso}(\mathcal{A})-\widehat{\tau}_{\acw}^{\lasso}(\tilde{\mathcal{A}})\}\rightarrow0$
when $N\rightarrow\infty$. 

\end{lemma}
Therefore, the Wald-type confidence interval can be constructed based
on $\widehat{\tau}_{\acw}^{\lasso}(\tilde{\mathcal{A}})$ given Lemmas
\ref{lem:acw_lass_fixed} and \ref{lem:acw_lass_selection}.

\begin{proof}[of Lemma \ref{lem:acw_lass_fixed}]
We can decompose the estimator $\widehat{\tau}_{\acw}^{\lasso}(\mathcal{A})$
into two parts as 
\[
\widehat{\tau}_{\acw}^{\lasso}(\mathcal{A})=N^{-1}\sum_{i\in\mathcal{R}\cup\mathcal{E}}\{\xi_{\acw,c}^{\lasso}(V_{i};\widehat{\mu}_{1},\widehat{\mu}_{0},\widehat{q},\widehat{r}_b)+\xi_{\acw,h}^{\lasso}(V_{i};\widehat{\mu}_{1},\widehat{\mu}_{0},\widehat{q},\widehat{r}_b)\},
\]
where
\begin{align*}
\xi_{\acw,c}^{\lasso}(V_{i};\widehat{\mu}_{1},\widehat{\mu}_{0},\widehat{q},\widehat{r}) & =\frac{R_{i}N}{N_{\mathcal{R}}}\left[\frac{A_{i}\{Y_{i}-\widehat{\mu}_{1}(X_{i})\}}{\pi_{A}(X_{i})}+\widehat{\mu}_{1}(X_{i})-\widehat{\mu}_{0}(X_{i})\right]\\
 & -\frac{R_{i}N}{N_{\mathcal{R}}}\frac{\widehat{q}(X_{i})}{\widehat{G}(X_{i})}(1-A_{i})\left\{ Y_{i}-\widehat{\mu}_{0}(X_{i})\right\} ,\\
\xi_{\acw,h}^{\lasso}(V_{i};\widehat{\mu}_{1},\widehat{\mu}_{0},\widehat{q},\widehat{r}) & =-\frac{(1-R_{i})N}{N_{\mathcal{R}}}\frac{\mathds{1}(\tilde{b}_{i}=0)\widehat{q}(X_{i})\widehat{r}_b(X_i)}{\widehat{G}(X_{i})}\left\{ Y_{i}-\widehat{\mu}_{0}(X_{i})\right\} ,
\end{align*}
where $\widehat{G}(X_{i})=\widehat{r}(X_{i})\widehat{P}(\tilde{b}_{i}=0\mid X, R=0)+\{1-\pi_{A}(X_{i})\}\widehat{q}(X_{i})$.

Asymptotically, $\widehat{\pr}\{\xi_{\acw,c}^{\lasso}(V_{i};\widehat{\mu}_{1},\widehat{\mu}_{0},\widehat{q},\widehat{r}_b)\}$
is consistent for $\tau$ under Assumption \ref{a:ign}. Denote the asymptotic limit of $\widehat{\pr}\{\xi_{\acw,h}^{\lasso}(V_{i};\widehat{\mu}_{1},\widehat{\mu}_{0},\widehat{q},\widehat{r}_b)\}$
by $\pr\{\xi_{\acw,h}^{\lasso}(V)\}$, where
\[
\xi_{\acw,h}^{\lasso}(V)=-\frac{1-R}{P(R=1)}\frac{\mathds{1}(b_{0}=0)q(X)r_b(X)}{G(X)}\left\{ Y-\mu_{0}(X)\right\}.
\]
Next, we can show the expectation of $\xi_{\acw,h}^{\lasso}(V_{i})$
is
\begin{align*}
 & \pr\left[\frac{1-R_{i}}{P(R_{i}=1)}\frac{\mathds{1}(b_{i,0}=0)q(X_{i})r_b(X_{i})}{G(X_{i})}\left\{ Y_{i}-\mu_{0}(X_{i})\right\} \right]\\
 & =\frac{1}{P(R_{i}=1)}\\
 & \times\pr\left[\frac{P(R_{i}=1\mid X_{i})\mathds{1}(b_{i,0}=0)r_b(X_{i})}{G(X_{i})}\left\{ \E(Y_{i}\mid X_{i},R_{i}=0,b_{i,0}=0)-\mu_{0}(X_{i})\right\} \right]\\
 & =0,
\end{align*}
where the last line holds because $\E(Y_{i}\mid X_{i},R_{i}=0,b_{i,0}=0)-\mu_{0}(X_{i})=0$
by definition. Therefore, $\widehat{\tau}_{\acw}^{\lasso}(\mathcal{A})$
is consistent for $\tau$ and its asymptotic normality follows by
standard M-estimation theory.
\end{proof}

\begin{proof}[of Lemma \ref{lem:acw_lass_selection}]
The difference between $\widehat{\tau}_{\acw}^{\lasso}(\mathcal{A})$
and $\widehat{\tau}_{\acw}^{\lasso}$ is 
\begin{align*}
\widehat{\tau}_{\acw}^{\lasso}(\mathcal{A})-\widehat{\tau}_{\acw}^{\lasso}(\tilde{\mathcal{A}}) & =N_{\mathcal{R}}^{-1}\sum_{i\in\mathcal{R}\cup\mathcal{E}}\frac{(1-R_{i})\{\mathds{1}(i\in\tilde{\mathcal{A}})-\mathds{1}(i\in\mathcal{A})\}\widehat{q}(X_{i})\widehat{r}}{\widehat{G}(X_{i})}\left\{ Y_{i}-\widehat{\mu}_{0}(X_{i})\right\} \\
 & =N_{\mathcal{R}}^{-1}\sum_{i\in\mathcal{A}}\frac{(1-R_{i})\{\mathds{1}(i\in\tilde{\mathcal{A}})-1\}\widehat{q}(X_{i})\widehat{r}}{\widehat{G}(X_{i})}\left\{ Y_{i}-\widehat{\mu}_{0}(X_{i})\right\} \\
 & +N_{\mathcal{R}}^{-1}\sum_{i\notin\mathcal{A}}\frac{(1-R_{i})\mathds{1}(i\in\tilde{\mathcal{A}})\widehat{q}(X_{i})\widehat{r}_b(X_i)}{\widehat{G}(X_{i})}\left\{ Y_{i}-\widehat{\mu}_{0}(X_{i})\right\} .
\end{align*}
By applying Lemma \ref{lem:selection-consistency}, we have $N_{\mathcal{R}}^{-1}\sum_{i\in\mathcal{A}}(1-R_{i})\{\mathds{1}(i\in\tilde{\mathcal{A}})-1\}h(V)\rightarrow0$
as $P(i\in\tilde{\mathcal{A}})\rightarrow1$ for $\forall i\in\mathcal{A}$
for any integrable function $h(\cdot)$. On the other hand, we have
$N_{\mathcal{R}}^{-1}\sum_{i\notin\mathcal{A}}(1-R_{i})\mathds{1}(i\in\tilde{\mathcal{A}})h(V)\rightarrow0$
as $P(i\in\tilde{\mathcal{A}})\rightarrow0$ for $\forall i\notin\mathcal{A}$
and any integrable function $h(\cdot)$. Therefore, we can obtain
the conclusion $N^{1/2}\{\widehat{\tau}_{\acw}^{\lasso}(\mathcal{A})-\widehat{\tau}_{\acw}^{\lasso}(\tilde{\mathcal{A}})\}\rightarrow0$
as desired.
\end{proof}

\subsection{Extension to Multiple External Controls \label{sec:Extension}}

In the previous sections, we illustrate the proposed data-adaptive
integrative estimator with only one external group $\rwd$.
In this section, we extend that to the case with multiple external
groups, denoted by $\rwd^{[1]},\ldots,\rwd^{[K]}$ with
size $N_{\rwd}^{[1]},\ldots,N_{\rwd}^{[K]}$, respectively.
The total sample size now is $N=N_{\rct}+\sum_{k=1}^{K}N_{\rwd}^{[k]}.$
Let $R_{i}^{[k]}$ be the data source  indicator for the external group
$\rwd^{[k]}$. An assumption similar to Assumption \ref{assum:exchange_delta}
is requested when integrating multiple external groups with the randomized placebo-controlled trial.
\begin{assumption}[Multiple external controls compatibility]\label{assum:exchange_delta-multiple}
For any $k=1,\ldots,K$, (i) 
$$\E\left\{ Y(0)\mid X=x,R_{i}^{[k]}=1\right\} =\E\{Y(0)\mid X=x,R_{i}^{[k]}=0\},$$ and (ii) $\pr(R^{[k]}=1\mid X=x)>0$
for all $x$ such that $\pr(X=x,R^{[k]}=1)>0$.
\end{assumption}

In a similar manner, the integrative augmented calibration weighting estimator $\widehat{\tau}_{\acw}^{[k]}$ for combining $\rwd^{[k]}$ with $\rct$ can be rewritten
by
\begin{align*}
\widehat{\tau}_{\acw}^{[k]} & =\frac{1}{N_{\rct}}\sum_{i\in\mathcal{\rct}\cup\rwd^{[k]}}R_{i}^{[k]}\left[\frac{A_{i}\{Y_{i}-\widehat{\mu}_{1}(X_{i})\}}{\pi_{A}(X_{i})}+\widehat{\mu}_{1}(X_{i})-\widehat{\mu}_{0}(X_{i})\right]\\
 & -\frac{1}{N_{\rct}}\sum_{i\in\mathcal{\rct}\cup\rwd^{[k]}}\frac{R_{i}^{[k]}(1-A_{i})\widehat{q}^{[k]}(X_{i}) + (1-R_{i}^{[k]})\widehat{q}^{[k]}(X_{i})\widehat{r}^{[k]}(X_{i})}{\widehat{r}^{[k]}(X_{i})+\{1-\pi_{A}(X_{i})\}\widehat{q}^{[k]}(X_{i})}\widehat{\epsilon}_{0,i},
\end{align*}
where $r^{[k]}(X)=\text{var}(Y\mid X,R^{[k]}=1, A=0)/\text{var}(Y\mid X,R^{[k]}=0)$, and $\widehat{q}^{[k]}(X)$ is the estimated
calibration weights balancing the covariate distribution in $\rwd^{[k]}$ to the target trial data. As illustrated previously, the conditional
mean exchangeability in Assumption \ref{assum:exchange_delta-multiple}
might be violated in practice. With slight modification, our proposed method is able to accommodate these potential violations presented in multiple external resources.
\begin{theorem}\label{thm:multiple-historical}
Let $\widehat{\tau}_{\acw}^{\lasso,[1:K]}=(\widehat{\tau}_{\acw}^{\lasso,[1]},\ldots,\widehat{\tau}_{\acw}^{\lasso,[K]})^{\T}$ be the concatenated data-adaptive integrative estimators for $K$ external groups, we have $N^{1/2}(\widehat{\tau}_{\acw}^{\lasso,[1:K]}-\tau)\rightarrow N(0,\Sigma_{\tau}^{\lasso})$. Thus, the final integrative estimator is $
\widehat{\tau}_{\acw}^{\final}=\widehat{d}^{\T}\widehat{\tau}_{\acw}^{\lasso,[1:K]}\rightarrow N(0,\widehat{d}^\T \Sigma_\tau^{\lasso}\widehat{d})$, where $\widehat{d}=\left\{ 1_{K}^{\T}(\Sigma_{\tau}^{\lasso})^{-1}1_{K}\right\} ^{-1}(\Sigma_{\tau}^{\lasso})^{-1}1_{K}$.\end{theorem}
The reason that we do not treat the multiple external groups as one entity
is due to the fact that different external resources might possess different covariate distributions since they can be collected by various registry databases. Therefore, it is optimal to calibrate each external group individually to the trial data for reaching stable weight estimation.

\begin{proof}[of Theorem \ref{thm:multiple-historical}]
    First, we present the detailed form of the data-adaptive integrative estimator $\widehat{\tau}_{\acw}^{\lasso, [k]}$ for the composite
observed data $\mathcal{\mathcal{R}}\cup\mathcal{E}^{[k]}$. For any $k=1,\cdots,K$, let $\widehat{{\xi}}^{[k]}$ be
the pseudo-observations of ${b}_{0}^{[k]}$, the penalized
least-square optimization can be formulated by
\[
\tilde{b}_{i}^{[k]}=\arg\min_{{b}}\left\{ (\widehat{{\xi}}^{[k]}-{b})^{\T}\widehat{\Sigma}_{\xi}^{-1}(\widehat{{\xi}}^{[k]}-{b})+\lambda_{N}\sum_{i\in\mathcal{E}^{[k]}}p(|b_{i}^{[k]}|)\right\} ,
\]
Next, the influence function for $\tau$ is modified based on ${b}_0^{[k]}$
\begin{align*}
 & \psi_{\tau,\acw}^{\lasso,[k]}(V;{\mu}_{1},{\mu}_{0},{q}^{[k]},{r}^{[k]},{\pi}_{b}^{[k]}, {b}_0^{[k]})\\
 & =\frac{R^{[k]}}{P(R^{[k]}=1)}\left[\{{\mu}_{1}(X)-{\mu}_{0}(X)-\tau\}+\frac{A}{\pi_{A}(X)}\{Y-{\mu}_{1}(X)\}\right]\\
 & -\frac{R^{[k]}(1-A)}{P(R^{[k]}=1)}\frac{{q}^{[k]}(X)}{{r}_b^{[k]}(X)P(b_0^{[k]}=0\mid X, R^{[k]}=1)+\{1-\pi_{A}(X)\}{q}^{[k]}(X)}\left\{ Y-{\mu}_{0}(X)\right\} \\
 & -\frac{(1-R^{[k]})\mathds{1}({b}_0^{[k]}=0)}{P(R^{[k]}=1)}\frac{{q}^{[k]}(X){r}_b^{[k]}(X)}{{r}_b^{[k]}(X)P(b_0^{[k]}=0\mid X, R^{[k]}=1)+\{1-\pi_{A}(X)\}{q}^{[k]}(X)}\{Y-{\mu}_{0}(X)\},
\end{align*}
where $r_b^{[k]}(X)=\text{var}(Y\mid X,R^{[k]}=1, A=0)/\text{var}(Y\mid X,R^{[k]}=0,b_{0}^{[k]}=0\}$. Then, the data-adaptive integrative estimator $\widehat{\tau}_{\acw}^{\lasso,[k]}$
for combing $\mathcal{R}$ and $\mathcal{E}^{[k]}$ is obtained by
solving the empirical analog of
\begin{align*}
\widehat{\tau}_{\acw}^{\lasso,[k]} & =\frac{1}{N_{\mathcal{R}}}\sum_{i\in\mathcal{\mathcal{R}}\cup\mathcal{E}^{[k]}}R_{i}^{[k]}\left[\frac{A_{i}\{Y_{i}-\widehat{\mu}_{1}(X_{i})\}}{\pi_{A}(X_{i})}+\widehat{\mu}_{1}(X_{i})-\widehat{\mu}_{0}(X_{i})\right]\\
 & -\frac{1}{N_{\mathcal{R}}}\sum_{i\in\mathcal{\mathcal{R}}\cup\mathcal{E}^{[k]}}\frac{R_{i}^{[k]}(1-A_{i})\widehat{q}^{[k]}(X_{i})}{\widehat{r}^{[k]}_b(X_i)
 \widehat{P}(\tilde{b}_i^{[k]}=0\mid X, R^{[k]}_i=1)+\{1-\pi_{A}(X_{i})\}\widehat{q}^{[k]}(X_{i})}\left\{ Y_{i}-\widehat{\mu}_{0}(X_{i})\right\} \\
 & -\frac{1}{N_{\mathcal{R}}}\sum_{i\in\mathcal{\mathcal{R}}\cup\mathcal{E}^{[k]}}\frac{(1-R_{i}^{[k]})\widehat{q}^{[k]}(X_{i})\widehat{r}^{[k]}_b(X_i)\mathds{1}(\tilde{b}_i^{[k]}=0)}{\widehat{r}^{[k]}_b(X_i)\widehat{P}(\tilde{b}_i^{[k]}=0\mid X, R_i^{[k]}=1)+\{1-\pi_{A}(X_{i})\}\widehat{q}^{[k]}(X_{i})}\left\{ Y_{i}-\widehat{\mu}_{0}(X_{i})\right\} ,
\end{align*}
and the asymptotic variance of $N^{1/2}(\widehat{\tau}_{\acw}^{\lasso,[1:K]}-\tau)$
is $\Sigma_{\tau}^{\lasso}=\E[\psi_{\tau,\acw}^{\lasso,[1:K]}(V)\{\psi_{\tau,\acw}^{\lasso,[1:K]}(V)\}^{\T}]$, which is based on the influence
functions of $\widehat{\tau}_{\acw}^{\lasso,[1:K]}$.
Next, the optimal integrative weight
$\widehat{{d}}$ to combine $\widehat{\tau}_{\acw}^{\lasso,[1:K]}$ are obtained by minimizing the post-integration variance:
\[
\min{d}^{\T}\Sigma_{\tau}^{\lasso}{d},\quad\text{subject to }{d}^{\T}1_{K}=1.
\]
By Lagrange multiplier method, we can show that $\widehat{{d}}=\left\{ 1_{K}^{\T}(\Sigma_{\tau}^{\lasso})^{-1}1_{K}\right\} ^{-1}(\Sigma_{\tau}^{\lasso})^{-1}1_{K}$
is the optimal combining weights. Hence, we have $\widehat{\tau}_{\acw}^{\final}=\widehat{{d}}^{\T}\widehat{\tau}_{\acw}^{\lasso,[1:K]}$
and $N^{1/2}(\widehat{\tau}_{\acw}^{\final}-\tau)\rightarrow N(0,\Sigma_{\tau}^{\final})$,
where $\Sigma_{\tau}^{\final}=\left\{ 1_{K}^{\T}(\Sigma_{\tau}^{\lasso})^{-1}1_{K}\right\} ^{-1}$.
\end{proof}


\section{Additional Simulation Results for Other Bias-generating Concerns \label{sec:More-Simulation-Results}}

\begin{figure}[!htbp]
    \centering
    \includegraphics[width=.85\linewidth]{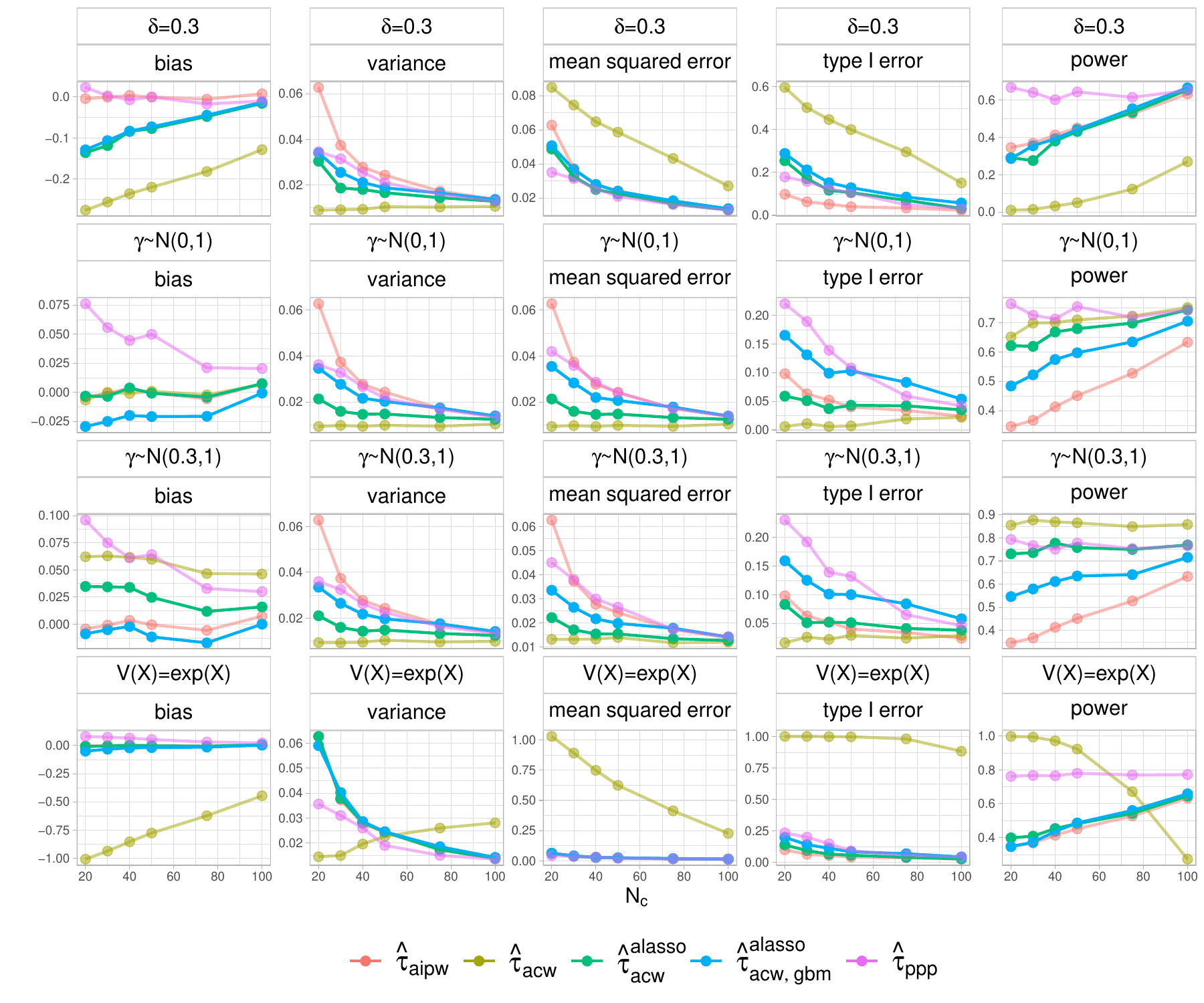}
\caption{\label{fig:additional_results} Additional simulation scenarios: (from
top to bottom) (i) lack of concurrency ($\delta_{T}=0.3$); (ii) measurement
error in covariate $X_{i,P}$ ($\gamma_{i}\sim N(0,1)$);
(iii) measurement error in covariate $X_{i,P}$ ($\gamma_{i}\sim N(0.3,1)$);
(iv) measurement error in outcome ($V(X)=\exp(X)$).}
\end{figure}

In this section, we investigate the performance of our proposed estimator of treatment effect in a placebo-control setting echoing the the Food and Drug Administration guidance for rare diseases, which outlines several
significant concerns regarding the use of external controls \citep{us2019rare}.
Specifically, the outcomes for $R_{i}=0$ are generated by: 
\begin{equation}
Y_{i}=V(\beta^{\T}X_{i}^{*}+\omega U_{i}\sigma_{Y}+\omega\sigma_{Y})+\delta_{T}T_{i}+\epsilon_{i},\quad\epsilon_{i}\sim N(0,\sigma_{Y}^{2}).\label{eq:RWD-Y-all}
\end{equation}
In particular, $\omega$ represents the level of confoundedness, which
gauges the association between the unmeasured confounder with the
selection propensity and the outcome. $\delta_{T}$ and $T_{i}$
constitute the effect of concurrency bias, in which $T_{i},i\in\mathcal{E}$
is simulated by taking values of $(0,1,2)$ with probability $1/3$
and $\delta_{T}$ represents the level of inconcurrency. For the baseline
covariate $X_{i}^{*}$ in the external control group, we have $X_{i}^{*}=(X_{i,1},\cdots,X_{i,d}^{*})$,
where $X_{i,d}^{*}=X_{i,d}+\gamma_{i}\sigma_{Y}$ and $\gamma_{i}$
represents the additional measurement error. $V(\cdot)$ represents
the outcome validity deviations and can be any non-linear function.
Table \ref{tab:Summary-of-simulated} summarizes all the bias-generating
issues raised by the Food and Drug Administration guidelines. To evaluate our dynamic borrowing under each bias-generating concern,
the corresponding assumption violation is evaluated separately in
Figure \ref{fig:additional_results}. Our proposed estimators $\widehat{\tau}_{\acw}^{\lasso}$
and $\widehat{\tau}_{\acw,\gbm}^{\lasso}$ both
achieve well-controlled mean squared error and improved power to detect meaningful treatment effects. Other similar findings can also be observed in the main paper.

\begin{table}[!htbp]
\centering
\caption{Summary of simulated scenarios with corresponding parameter values}
\vspace{0.5em}{%
\begin{tabular}{ll}
\toprule 
Scenario & {Values}\\
\midrule 
Unmeasured confounding & {$\omega=(0,0.3)$}\\
Lack of concurrency & {$\delta_{T}=(0,0.3)$}\\
Measurement errors & {$\gamma_{i}=\{0, N(0,1), N(0.3,1)\}$}\\
Outcome validity & $V(X)=\{X,\exp(X)\}$\\
Treatment effect & $\E(\alpha^{\T}X)=0$, $\E(\alpha^{\T}X)=0.3$\\
\bottomrule
\end{tabular}}\label{tab:Summary-of-simulated}
\end{table}

\section{Additional Real-data Application \label{sec:application-II}}

In this section, we study the effect of solanezumab, an anti-amyloid Beta monoclonal antibody, on the progression of mild Alzheimer’s Disease. The goal is to test whether intravenous infusion of solanezumab can slow the rate of cognitive decline and functional impairment compared to placebo. To do so, the control arm of a randomized placebo-controlled trial is augmented with a real-world observational study. EXPEDITION2 (https://clinicaltrials.gov/ct2/show/study/NCT00905372) is an 18-month multi-center, double-blind, phase III randomized
clinical trial with patients randomized 1:1 to receive solanezumab or placebo, which is considered the gold-standard randomized controlled trial. \textsc{geras-eu} \citep{wimo2013geras} is an 18-month, prospective observational study of patients
with Alzheimer’s Disease conducted across three European (EU) countries (France, Germany,
and the United Kingdom). 

\begin{figure}[htbp]
    \centering
    \includegraphics[width = .8\linewidth]{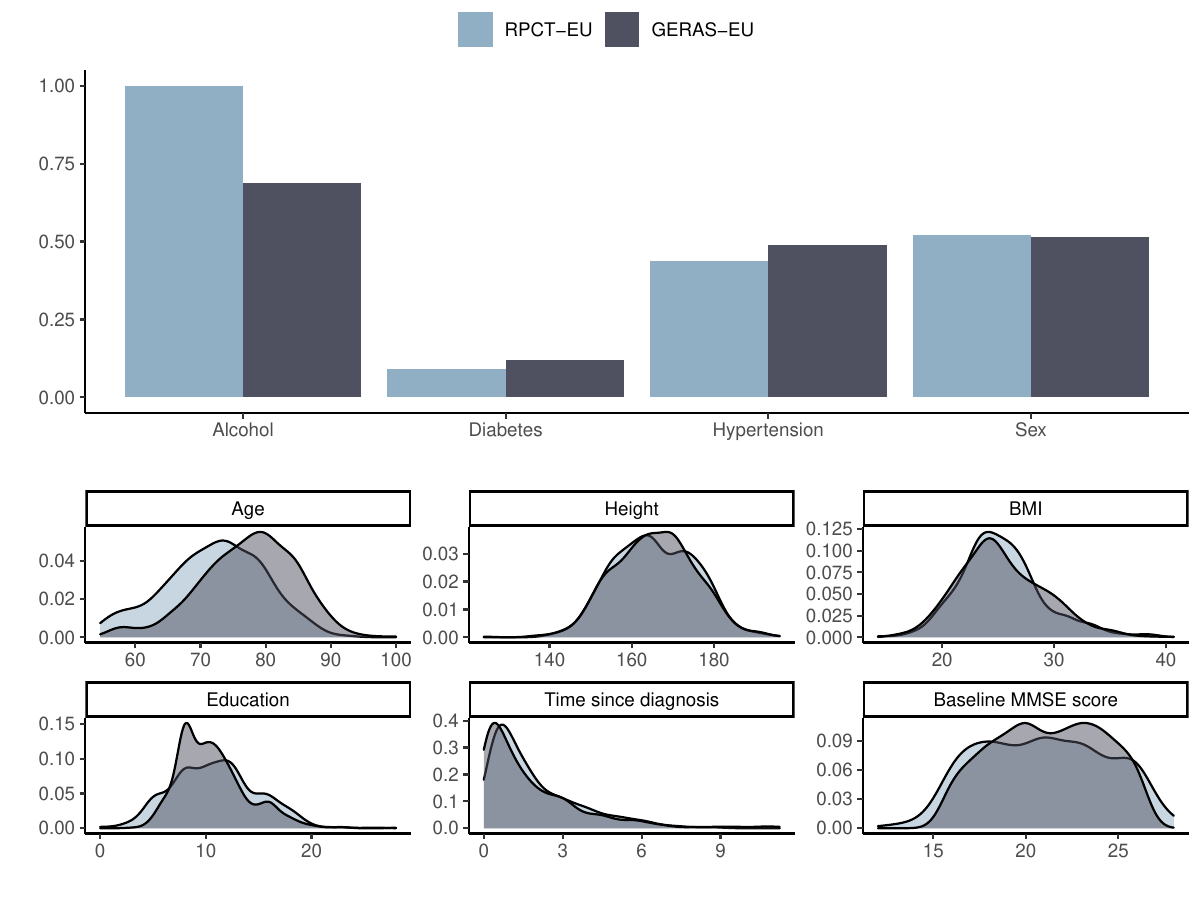}
    \caption{Distributions of age, sex, height, BMI, years of education, time since diagnosis of Alzheimer’s Disease, Mini-Mental State Examination score at baseline, alcohol consumer, hypertension, and diabetes in for \textsc{rpct-eu} and \textsc{geras-eu}.}
    \label{fig:AD_EDA}
\end{figure}

As noted by \citet{henley2015alzheimer} and \citet{grill2015comparing},
the multi-center randomized controlled trial may be subject to population heterogeneity stemming
from country-wise variations in metric standards and healthcare systems.
To facilitate comparisons across more homogeneous patient populations between the data sources, we select patients from the European populations (i.e.,
France, Germany, Italy, Poland, Spain, Sweden, and the United Kingdom) in EXPEDITION2, which will subsequently be denoted as \textsc{rpct-eu}. Next, similar to the analyses
in \citet{siemers2016phase} and \citet{reed2018representativeness},
we compare the development of Alzheimer’s Disease in a group of patients with probable
mild Alzheimer’s Disease dementia, pre-specified as those with a baseline Mini-Mental
State Examination score of 20-26 \citep{folstein1992practical}.
In \textsc{rpct-eu}, this results in $316$ eligible patients, with $154$ randomly assigned
to solanezumab and $162$ randomly assigned to the placebo. The
\textsc{geras-eu} database includes $578$ placebo patients.
Table \ref{tab:covariate-info} presents the descriptive statistics
of several demographic and baseline clinical characteristics contained in
both studies. A visual illustration of these study-specific covariate distributions at baseline is presented in Figure
\ref{fig:AD_EDA}. All continuous covariates are standardized
before being used in our adaptive-lasso framework. 

\begin{table}[htbp]
\centering
\caption{Patient baseline characteristics; data
is presented as means (standard deviations) for
continuous covariates and percentages for categorical covariates}
\vspace{0.5em}{%
\begin{tabular}{lll}
\toprule
Characteristic & \textsc{rpct-eu} (316) & \textsc{geras-eu} (578)\tabularnewline
\midrule
Age & $72.4(7.5)$ & $77.3(6.8)$\tabularnewline
Sex (0-Male, 1-Female) & $1(51.9\%)$ & $1(50.7\%)$\tabularnewline
Height & $166.3(9.6)$ & $165.9(9.8)$\tabularnewline
BMI & $25.3(3.6)$ & $25.4(3.9)$\tabularnewline
Education (years) & $11.4(4.0)$ & $10.8(3.2)$\tabularnewline
Time since diagnosis of AD (years) & $1.7(1.5)$ & $1.6(1.9)$\tabularnewline
Baseline MMSE score & $22.5(2.8)$ & $22.8(2.0)$\tabularnewline
Alcohol consumer (0-No, 1-Yes) & $1(47.2\%)$ & $1(67.8\%)$\tabularnewline
Hypertension (0-No, 1-Yes) & $1(44.9\%)$ & $1(51.7\%)$\tabularnewline
Diabetes (0-No, 1-Yes) & $1(7.6\%)$ & $1(12.8\%)$\tabularnewline
\bottomrule
\end{tabular}}\label{tab:covariate-info}
\end{table}

Treatment effects are evaluated between patients treated with solanezumab
from the \textsc{rpct-eu} study and a hybrid control group consisting of the
placebo group from \textsc{rpct-eu} and the entire \textsc{geras-eu} cohort. The progression of
Alzheimer’s Disease is quantified by changes in patients' cognition, functional
ability, neuropsychiatric symptoms, and health-related quality of
life (HRQoL) over 18 months. Cognition is assessed by the 14-item
cognitive subscale of the Alzheimer's Disease Assessment Scale (ADAS-Cog
14 \citep{rosen1984new,mohs1997development}), where poor cognition corresponds
to higher ADAS-Cog 14 scores. Functional ability is measured by the instrumental
activities of daily living (iADL) score \citep{galasko1997inventory}.
Greater impaired functioning is reflected by lower iADL scores. Neuropsychiatric
symptoms are examined according to the Neuropsychiatric Inventory
(NPI \citep{cummings1994neuropsychiatric}) item. The NPI total score ranges
from $0$ to $144$, with higher scores indicating greater distress.
Lastly, HRQol is quantified by the EuroQol-5 Dimensions questionnaire
(EQ-5D \citep{kind1996euroqol}) where a lower score
corresponds to reduced HRQoL in the measured patients. We assume a
linear heterogeneous treatment effect function with the characteristics
in Table \ref{tab:covariate-info} as the treatment effect modifiers.
The same set of estimators in the simulation study are considered
and the results are presented in Table \ref{tab:est_res-alzheimer}.
Due to the limited sample size of \textsc{rpct-eu}, $\widehat{\tau}_{\aipw}$
is not able to detect significant treatment effects for any of the outcomes. By leveraging
the external observational study, the integrative augmented calibration weighting estimator $\widehat{\tau}_{\acw}$
modifies the point estimates and reduces standard error for estimating the treatment differences of solanezumab in terms of iADL and NPI compared to placebo, which collectively leads to statistical significance. However, the findings based on $\widehat{\tau}_{\acw}$
may be subject to possible biases as they are greatly different from $\widehat{\tau}_{\aipw}$. Lastly, the data-adaptive integrative
estimator $\widehat{\tau}_{\acw}^{\lasso}$ selectively incorporates
comparable patients only and produces point estimates that are closer
to $\widehat{\tau}_{\aipw}$ than to $\widehat{\tau}_{\acw}$, but
with smaller standard errors than $\widehat{\tau}_{\aipw}$. As a
result, the proposed estimator is able to improve treatment effect estimation
in terms of ADAS-Cog 14 and yield statistically significant treatment effect.

\begin{table}[!ht]
\centering
\caption{Point estimates, standard errors, and
95\% confidence intervals of the treatment effect of solanezumab against
the placebo regarding different outcomes based on the \textsc{rpct-eu} and \textsc{geras-eu}
studies}
\vspace{0.5em}{%
\begin{tabular}{lccccc}
\toprule
Outcome &  & $\widehat{\tau}_{\aipw}$ & $\widehat{\tau}_{\acw}$ & $\widehat{\tau}_{\acw}^{\lasso}$ & $\widehat{\tau}_{\ppp}$\tabularnewline
\midrule
ADAS-Cog 14 & Est. & -2.74 & -0.19 & -2.74 & -2.72\tabularnewline
 & S.E. & 1.52 & 1.20 & 1.30 & 1.56\tabularnewline
 & C.I. & (-5.71, 0.23) & (-2.54, 2.17) & (-5.28, -0.20) & (-5.77, 0.33)\tabularnewline
 &  &  &  &  & \tabularnewline
iADL & Est. & 1.71 & -8.57 & 1.71 & 1.77\tabularnewline
 & S.E. & 1.68 & 1.43 & 1.52 & 1.73\tabularnewline
 & C.I. & (-1.58, 5.01) & (-11.38, -5.77) & (-1.27, 4.69) & (-1.63, 5.17)\tabularnewline
 &  &  &  &  & \tabularnewline
NPI & Est. & -2.35 & -3.50 & -2.35 & -2.29\tabularnewline
 & S.E. & 1.75 & 1.41 & 1.41 & 1.74\tabularnewline
 & C.I. & (-5.78, 1.08) & (-6.26, -0.73) & (-5.12, 0.42) & (-5.70, 1.11)\tabularnewline
 &  &  &  &  & \tabularnewline
EQ-5D & Est. & 0.94 & -0.41 & 0.94 & 0.93\tabularnewline
 & S.E. & 2.39 & 1.99 & 2.09 & 2.43\tabularnewline
 & C.I. & (-3.74,5.63) & (-4.31,3.49) & (-3.16,5.04) & (-3.83,5.70)\tabularnewline
\bottomrule
\end{tabular}}
\label{tab:est_res-alzheimer}
\end{table}

\begin{figure}[htbp]
    \centering
    \includegraphics[width = .8\linewidth]{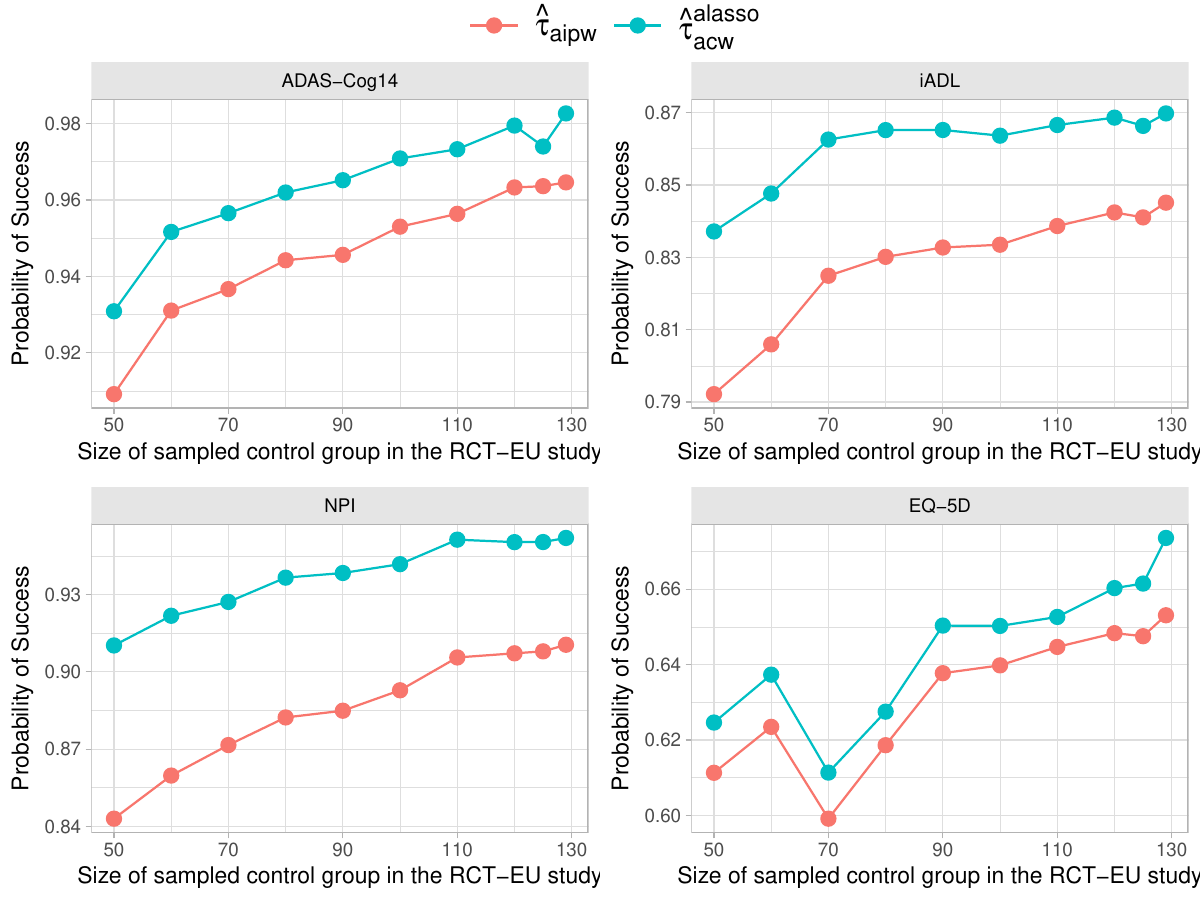}
    \caption{Probability of success by estimators
$\widehat{\tau}_{\aipw}$ and $\widehat{\tau}_{\acw}^{\lasso}$ with
varying control group size of the \textsc{rpct-eu} study.}
\label{fig:sub-sampling_figs}
\end{figure}

Next, we adopt the same sub-sampling strategy used in Section \ref{sec:application} to highlight the benefits of our proposed method.
Particularly, we randomly select $N_{c}^{s}$ patients from the placebo
group of the \textsc{rpct-eu} study to create $100$ sub-samples, where $N_{c}^{s}=50,60,\cdots,120,125,130$.
Figure \ref{fig:sub-sampling_figs} depicts the averaged probability of
success, computed by the likelihood of successfully detecting $\tau<0$
for ADAS-Cog 14 and NPI, but $\tau>0$ for iADL and EQ-5D. This experiment
highlights the merits of our data-adaptive borrowing framework as we can achieve higher success probabilities compared with the benchmark trial-only estimator in all aspects of assessing mild Alzheimer’s Disease progression. Thus, our dynamic borrowing framework can effectively reduce the expenditure of the randomized clinical trial by recruiting fewer patients for the concurrent placebo arm and utilizing the real-world external controls as supplements to maintain a similar trial power.

\end{document}